\documentclass{ieeeaccess}

\usepackage{graphicx}
\usepackage{amsmath,amssymb,amsfonts}

\usepackage{enumitem}
\usepackage{cite}
\usepackage{textcomp}
\usepackage{pifont}
\usepackage[export]{adjustbox}
\usepackage[labelformat=simple]{subfig}

\usepackage{hyperref}

\newcommand{\myparatight}[1]{\vspace{0.5ex}\noindent\textbf{#1~}}
\usepackage{stfloats}
\usepackage{makecell}
\usepackage{soul}
\usepackage[capitalize]{cleveref}
\crefname{table}{Tab.}{Tabs.}
\usepackage[font=small]{caption}
\definecolor{lightgray}{gray}{0.9}
\definecolor{lightblue}{rgb}{0.9,0.9,1}
\definecolor{red}{rgb}{1,0,0}

\newcommand{\name}{AgentOptics}

\def\BibTeX{{\rm B\kern-.05em{\sc i\kern-.025em b}\kern-.08em
    T\kern-.1667em\lower.7ex\hbox{E}\kern-.125emX}}

\begin{document}
\history{Date of publication xxxx 00, 0000, date of current version May 15, 2026.}
\doi{10.1109/ACCESS.2017.DOI}

\title{Agentic AI for Scalable and Robust Optical Systems Control}




\author{\uppercase{Zehao Wang}\authorrefmark{1,*}, \uppercase{Mingzhe Han}\authorrefmark{1,*},
\uppercase{Wei Cheng}\authorrefmark{1}, \uppercase{Yue-Kai Huang}\authorrefmark{3},
\uppercase{Philip N. Ji}\authorrefmark{3}, \uppercase{Denton Wu}\authorrefmark{2},
\uppercase{Mahdi Safari}\authorrefmark{6}, \uppercase{Flemming Holtorf}\authorrefmark{6},
\uppercase{Kenaish AlQubaisi}\authorrefmark{7}, \\ 
\uppercase{Danyang Zhuo}\authorrefmark{1},
\uppercase{Yiran Chen}\authorrefmark{1}, \uppercase{Ting Wang}\authorrefmark{3},
\uppercase{Norbert M. Linke}\authorrefmark{2}\authorrefmark{4}\authorrefmark{5},
\uppercase{Dirk Englund}\authorrefmark{7}, and \uppercase{Tingjun Chen}\authorrefmark{1}}
\address[1]{Department of Electrical and Computer Engineering, Duke University, Durham, NC 27708 USA}
\address[2]{Duke Quantum Center and Department of Physics, Duke University, Durham, NC 27708 USA}
\address[3]{NEC Laboratories America, Princeton, NJ 08540 USA}
\address[4]{Joint Quantum Institute and Department of Physics, University of Maryland, College Park, MD 20742 USA}
\address[5]{National Quantum Laboratory (QLab), University of Maryland, College Park, MD 20742 USA}
\address[6]{Axiomatic AI, Cambridge, MA 02139 USA}
\address[7]{Research Laboratory of Electronics, Massachusetts Institute of Technology, Cambridge, MA 02139 USA}
\tfootnote{This work was supported in part by NSF under grants EEC-1941583, CNS-2112562, OIA-2134891, CNS-2211944, PHY-2325080, CNS-2330333, CNS-2443137, CNS-2450567, and OAC-2503010, and ARO under grant W911NF-25-1-0241.
We also acknowledge funding from Duke University under the Beyond-the-Horizon and DST-Launch initiatives.}
\markboth
{Wang \headeretal: Agentic AI for Scalable and Robust Optical Systems Control}
{Wang \headeretal: Agentic AI for Scalable and Robust Optical Systems Control}
\corresp{*These authors contributed equally to this work. Corresponding authors: Zehao Wang and Mingzhe Han (zehao.w@duke.edu, mingzhe.han@duke.edu).}

\begin{abstract}
We present {\name}, a language-driven orchestration layer for high-fidelity, autonomous optical system control built upon the Model Context Protocol (MCP).
{\name} interprets natural language tasks and executes standardized, schema-validated actions on heterogeneous optical devices through a structured tool abstraction layer.
We implement 64 standardized MCP tools spanning eight representative optical devices and construct a comprehensive 410-task benchmark to evaluate the performance of {\name} across request understanding, role-dependent responses, multi-step coordination, robustness to linguistic variation, and error-handling capability.
We evaluate two deployment configurations, integrating either commercial online large language models (LLMs) or locally deployed open-source LLMs, and compare against LLM-based code generation baselines.
Experimental results demonstrate that {\name} achieves average task success rates from 65.4\% to 98.8\% across the evaluated local and online models, significantly outperforming code generation approaches with up to only 55\% success rate.
We further validate the broader applicability of {\name} through five representative case studies covering dense wavelength division multiplexing (DWDM) link provisioning, analog radio-over-fiber (ARoF) link characterization, multi-span channel optimization, polarization stabilization, and distributed acoustic sensing (DAS)-based fiber monitoring with LLM-assisted event interpretation. We also demonstrate an LLM-assisted workflow that generates MCP-compatible tool interfaces from legacy device-control code, illustrating a scalable path for onboarding additional optical device capabilities.
These results demonstrate that {\name} provides a scalable and robust interface for natural-language-driven control and orchestration of heterogeneous optical devices and systems.
\end{abstract}

\titlepgskip=-15pt


\maketitle

\section{Introduction}


Optical networks form the backbone of modern Internet infrastructure, interconnecting data centers, metro and long-haul transport systems, wireless front/backhaul, and emerging quantum networks~\cite{nishizawa2024semi,nishizawa2024fast,sasai2025optical,wang2025toward,ferrari2020gnpy,yu2019cosmos,simon2017towards,sevincer2013lightnets}.
As optical systems scale in both heterogeneity and performance, incorporating reconfigurable optical add-drop multiplexers (ROADMs), coherent pluggable transceivers (TRXs), radio-over-fiber (RoF) links, fiber sensing, and quantum photonic hardware, the operational complexity of device configuration, monitoring, and optimization increases significantly.
Achieving high fidelity, reliability, and efficiency in these systems requires coordinated control across heterogeneous devices, accommodation of vendor-specific management interfaces, and closed-loop telemetry and adaptation mechanisms.

Software-defined networking (SDN) has been widely adopted in optical networks to decouple the control plane from the data transmission plane, enabling centralized control, programmability, and automated provisioning of optical resources~\cite{ferrari2020gnpy,borraccini2021autonomous,raychaudhuri2020challenge,chen2022software}.
By abstracting network control, SDN improves operational efficiency and flexibility in managing complex optical infrastructures.
Initiatives such as OpenROADM~\cite{casellas2024openROADM} also aim to address this challenge by defining common data models and standardized control interfaces across multi-vendor optical transport equipment, improving system interoperability and manageability.

\begin{figure*}[!t]
    \centering
    \includegraphics[width=0.98\textwidth]{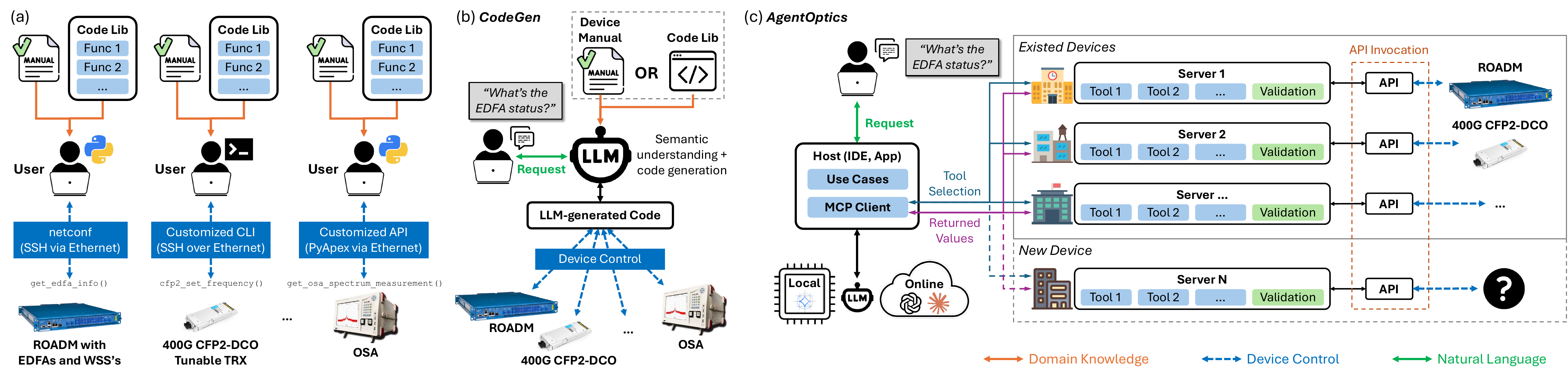}
    \caption{Optical device control using ROADM, 400\thinspace{GbE} CFP2-DCO, and OSA as examples:
    (a) Traditional control requires device-specific manuals, custom scripts, and protocol handling.
    (b) LLM-based control interprets natural-language prompts to generate control code, reducing manual scripting.
    (c) The proposed {\name} framework standardizes control via a unified tool layer, where the MCP client maps prompts to device APIs through tool selection, enabling remote distributed device control and scalable integration of new devices.}
    \label{fig:overview-diagram}
    \vspace{-3.0mm}
\end{figure*}

However, existing SDN solutions face limitations in multi-vendor environments due to inconsistent support for standardized interfaces and vendor-specific extensions.
In addition, limited abstraction of physical-layer behaviors and device capabilities increases system complexity and poses challenges for operators seeking scalable and intuitive network control. 
In current practice, SDN low-level control layers often rely on human-written scripts derived from vendor manuals or implemented through software development kits (SDKs) and command-line interface (CLI) tools, as shown in Fig.~\ref{fig:overview-diagram}(a).
Developers and users must translate high-level experimental goals (e.g., ``\textit{get optical power spectrum}'') into explicit sequences of function calls or protocol commands, e.g., via Network Configuration Protocol (NETCONF), or vendor-provided Python application programming interfaces (APIs), manually specify execution order, establish device connections, validate parameter constraints, and parse raw device responses.
While this workflow offers precise control, it requires substantial engineering work and device-specific expertise, resulting in significant onboarding effort and limited portability across heterogeneous equipment. 

Recent advances in large language models (LLMs) have enabled intuitive, natural language-based interaction with complex systems, transforming how users control, configure, and automate technical workflows~\cite{hou2024large}.
The primary advantage of LLM-based agents is not merely improved natural language understanding, but the ability to translate high-level design objectives into structured, multi-step execution workflows, augmented with reasoning and decision-making capabilities.
However, enabling reliable interaction between LLMs and heterogeneous technical systems requires more than improved reasoning capability: it demands a standardized and structured interface for tool discovery, invocation, and response handling.
The Model Context Protocol (MCP)~\cite{modelcontextprotocol2025intro} addresses this need by defining a formal client-server architecture that connects LLM hosts to external tools and services through well-defined schemas and execution semantics.
By decoupling reasoning from execution, MCP provides a unifying protocol for intelligent, model-driven interaction with diverse and distributed technical environments.

LLM-based agents have already been widely adopted in domains such as software engineering~\cite{yang2024sweagent} (e.g., repository management, issue tracking), personal productivity~\cite{wijerathne2025scheduleme} (e.g., calendar scheduling, daily task automation), and data analytics~\cite{wang2025llmagentdrivendataanalysis} (e.g., structured knowledge base querying).
More recently, tool-augmented or MCP-enabled LLM systems have been expanded and applied to a broader range of applications, including communication and edge computing systems~\cite{xu2024llm6g}, Internet-of-Things networks~\cite{iotmcp}, electronic design automation~\cite{mcp4eda}, and photonic integrated circuit design~\cite{sharma2025phido}.
These developments demonstrate the growing role of LLM agents as orchestration layers that bridge natural language intent and structured system interfaces.

Within networking and optical systems, LLM-based agents have been explored for monitoring, diagnosis, control, and performance optimization~\cite{wang2024alarmGPT,pang2024llama2loganalysis,zhang2025autolight,liu2025first}.
In most approaches involving device control, the agent translates user requests, often combined with device manuals, schema definitions, or reference code, into executable control scripts or structured SDN API calls.
This workflow, illustrated in Fig.~\ref{fig:overview-diagram}(b), relies on the LLM to synthesize control logic that directly interfaces with device APIs or controller northbound interfaces.
While such code-generation or API-synthesis approaches reduce manual scripting effort, they remain tightly coupled to textual reasoning and prompt conditioning.
In contrast, the MCP-based design in {\name} separates device-side execution from direct LLM-generated control logic through decoupled client and server components, motivating a more structured interface between reasoning and physical device actuation.
At the level of significant challenges persist, particularly in ensuring high-fidelity tool invocation, strict parameter validation, and robust handling of dynamic or ambiguous user inputs across heterogeneous devices.
These limitations motivate the need for a more structured abstraction layer between LLM reasoning and physical device execution. 


In this paper, we present {\name}, an MCP-based orchestration framework for autonomous, scalable, and high-fidelity optical device control.
As shown in Fig.~\ref{fig:overview-diagram}(c), {\name} serves as a language-driven orchestration layer that provides a standardized interface between LLM-based reasoning and heterogeneous optical device APIs, translating user natural language inputs into protocol-compliant operations by invoking corresponding MCP tools across heterogeneous optical devices through validated API calls.
We implement 64 standardized MCP tools spanning eight representative optical devices (see Table~\ref{tab:mcp-tool-list}), encapsulating common device operations as deterministic, executable primitives.
{\name} contributes a complete MCP-based control framework for optical systems: it defines how heterogeneous optical-device APIs are exposed as validated, schema-guided tools, how LLMs select and coordinate these tools across devices, and how the resulting workflow is evaluated on real hardware under multi-step, multi-device, and error-prone user requests.
By exposing device operations through structured MCP-based tool schemas and leveraging LLMs, {\name} enables dynamic workflow orchestration without requiring task-specific code generation.

We systematically evaluate {\name} through a comprehensive benchmark on real optical hardware, demonstrate its system-level orchestration capabilities through five representative runtime case studies, and further illustrate its extensibility through an LLM-assisted MCP tool-generation workflow.
In particular, we construct a benchmark consisting of 410 tasks executed on real hardware and systems, covering single-, dual-, and triple-action invocations across multiple devices. 
The benchmark also includes five representative task variants, including \textit{paraphrasing}, \textit{non-sequitur}, \textit{error}, \textit{roles}, and \textit{chain}, designed to emulate realistic and diverse user inputs to assess the robustness and fidelity of {\name} under dynamic interaction scenarios. 
We implement {\name} using four representative commercial online LLMs from the GPT and Claude families, as well as five locally hosted Gemma and Llama models. For comparison, we establish an LLM-based code generation (CodeGen) baseline with two variants:
(\textit{i}) online LLM-based optical device control via direct code generation, conditioned on either device manuals or reference control code, and
(\textit{ii}) a locally deployed LLM fine-tuned on optical device control code using low-rank adaptation (LoRA).

Overall, {\name} achieves an average success rate of 98.8\% with online LLMs and 65.4\% across the five locally hosted LLMs on the 410-task benchmark.
In contrast, the CodeGen baseline attains average success rates of up to only 50.0\%.
{\name} enables practical natural language control of optical devices, achieving 98.1\% to 99.3\% success rates with online LLMs at a token cost ranging from \$0.004 to \$0.15 per task, while the evaluated locally hosted Gemma and Llama models achieve 7.1\% to 84.4\% success rates at near-zero API cost per task.

We further demonstrate six representative case studies to illustrate the capabilities and extensibility of {\name} through:
(\textit{i}) For rapid device onboarding, we demonstrate an LLM-assisted workflow that converts legacy optical spectrum analyzer (OSA) control code into MCP-compatible tool interfaces and validates the generated server on physical hardware.
(\textit{ii}) In a ROADM-based dense wavelength-division multiplexing (DWDM) network, {\name} provisions a coherent 400\thinspace{Gigabit Ethernet (GbE)} signal alongside an analog radio-over-fiber (ARoF) signal and performs coordinated multi-device monitoring of link performance metrics, including optical signal-to-noise ratio (OSNR) and error vector magnitude (EVM).
(\textit{iii}) For a wideband ARoF link carrying 5G fronthaul traffic, {\name} autonomously characterizes system performance and optimizes the ARoF transmitter bias voltage to enhance wireless transmission quality.
(\textit{iv}) In a two-span link with co-propagating comb channels, {\name} provisions additional 400\thinspace{GbE} channels and autonomously adjusts launch power to minimize the pre–forward error correction (pre-FEC) bit error rate (BER).
(\textit{v}) For polarization-sensitive links, {\name} enables closed-loop polarization stabilization through orchestrated measurement and actuation, maintaining convergence despite intentional fiber perturbations.
(\textit{vi}) For distributed acoustic sensing (DAS)-based fiber sensing systems, {\name} automates sensing operations, data acquisition, and LLM reasoning to identify potential fiber cut events. 
{\name} implementation and benchmark are open-sourced at~\cite{github_agentoptics}.

This paper is organized as follows.
We review the background and related work in Section~\ref{sec:related_work}, and describe the proposed system architecture and implementation in Section~\ref{sec:methods}.
Sections~\ref{sec:evaluation} and~\ref{sec:results} present the benchmarking methodology and experimental results.
We demonstrate practical capabilities of {\name} through five runtime case studies and one LLM-assisted tool-generation case study in Section~\ref{sec:casestudies}, and conclude in Section~\ref{sec:conclusions}.

\section{Related Work}
\label{sec:related_work}

\subsection{Agentic AI Frameworks and Applications}
Unlike traditional AI systems that generate single-shot text responses, agentic AI extends LLMs to perform goal setting, multi-step planning, external tool interaction, and feedback-driven decision-making.
A defining feature of such agents is their ability to act on external systems, typically through structured tool function calling.
There are multiple methods through which an LLM may invoke a tool function: the tool name and definition may be
\textit{(i)} implicitly acquired during pre-training~\cite{schick2023toolformer}, which requires massive training data;
\textit{(ii)} provided as part of the input prompt and executed via an external controller~\cite{yao2022react}, where the context length increases linearly with tool count;
\textit{(iii)} accessed through a specific protocol such as MCP~\cite{modelcontextprotocol2025intro}, which provides a standardized schema but adds protocol overhead; or
\textit{(iv)} implemented via program-aided language (PAL) models~\cite{gao2023pal}, where the model directly generates executable control code, which offers flexibility but lacks safety validation. 
General-purpose agent orchestration frameworks such as LangChain/LangGraph~\cite{langgraph_docs}, AutoGen~\cite{wu2024autogen}, and CrewAI~\cite{crewai_docs} provide high-level abstractions for tool use, planning, memory, and multi-agent coordination. However, these frameworks do not by themselves define validated physical-device control interfaces for optical hardware. To operate real optical devices, they still require device APIs, tool schemas, parameter validation, execution guards, and hardware-specific backend logic. Therefore, our comparison focuses on the device-control workflow that is most directly comparable in this setting: direct LLM-based code generation from manuals or reference code versus structured MCP-based tool invocation over validated backend device-control functions.

These methods enable LLMs to interact with a wide range of real-world applications. 
For example, HuggingGPT~\cite{shen2023hugginggpt} is an early example that uses an LLM as a controller to route user requests to specialized expert models and aggregate their output into a comprehensive response. 
SWE-agent~\cite{yang2024sweagent} demonstrates a repository-level automation agent for software engineering. 
IoT-MCP~\cite{iotmcp} bridges LLMs and heterogeneous devices for Internet of Things (IoT) system development.
In scientific reasoning and verification, ax-Prover~\cite{breen2025axprover} shows the agent's capability in theorem proving for mathematics and quantum physics.
Similarly, physics Supernova~\cite{qiu2025physics} demonstrates near-gold medal performance on International Physics Olympiad problems.
Seed-Prover~\cite{chen2025seed} reaches undergraduate- to PhD-level mathematics capability.
In addition, a multi-agent framework achieves single-device design~\cite{sharma2025phido}.
In the broader networking area, agentic pipelines have been explored for intent-based infrastructure and service orchestration~\cite{brodimas2025agentic}, as well as wireless and O-RAN management~\cite{wu2025oranagent}. 
Beyond networking, LLM-enabled cyber-physical systems have also been explored across broader engineering domains, where LLMs can serve as assistants for perception and interaction or as high-level components for reasoning and planning~\cite{xu2024llmenabled}. These systems share the broader theme of agentic reasoning over evolving physical environments, while {\name} focuses specifically on protocol-compliant control and orchestration of heterogeneous devices and network systems through validated MCP tool interfaces.
\subsection{Agentic AI in Optical Network Monitoring and Control}

In optical networks, recent LLM-based systems began to couple language-guided decision making with operational interfaces, enabling workflows that interpret telemetry and logs, coordinate control-plane actions, and verify outcomes through feedback from the network and devices.

\myparatight{Agentic optical network diagnosis and monitoring.}
Several studies have applied LLM agents to optical network diagnosis and monitoring tasks, primarily focusing on analytical reasoning and decision support.
In~\cite{zhang2025llmagent}, a GPT-4-powered agent was proposed to support autonomous optical network management, such as quality of transmission (QoT) estimation, performance analysis, optimization, and calibration.
AlarmGPT~\cite{wang2024alarmGPT} is a LangChain-based tool-augmented workflow that automates optical transport networks for alarm interpretation, compression, prioritization, and diagnosis. 
\cite{pang2024llama2loganalysis} presented an instruction-tuned LLM for field-collected optical network log parsing, anomaly detection and classification, and report generation. 

\myparatight{Agentic optical network control.}
Recent works have also explored LLM-based automation for optical network control~\cite{zhang2026aiagent}.
These approaches typically employ external grammars to convert natural language outputs into valid executable instructions, incorporate device API descriptions through prompt engineering, or fine-tune models to directly generate structured function calls.
For example,~\cite{dicicco2024open} proposed an LLM-driven pipeline that leverages formal grammars to constrain the LLM output to valid JSON-formatted device control instructions for SDN configuration.
Another approach to instruction generation is embedding device API specifications directly within prompts, and it was demonstrated in~\cite{liu2025first} that amplifier gain optimization can be performed via SDN API in prompts. 
Similarly, AutoLight presented in~\cite{zhang2025autolight} is a multi-agent framework for distributed AI training using optical communications API as LLM input reference.
Related LLM-based approaches have also been explored for multivendor transport network configuration management~\cite{wang2025llmcentric} and expertise-guided optical power optimization in field-deployed networks~\cite{qiu2025expertiseguided}.
Compared with these approaches, {\name} keeps device-specific execution within server-side tools while reusing the same client-side orchestration across heterogeneous optical devices, rather than relying on the LLM to directly generate device-level control code or controller API calls. This decoupled client–server design offers greater safety, policy enforcement, and flexibility for remote optical device operation.
Finally, smaller LLMs fine-tuned on network-specific control instructions have been used to directly generate executable commands~\cite{sun2025automation}.

However, existing LLM-based autonomous device and network control methods introduce three major limitations.
First, these approaches assume the existence of a mature SDN infrastructure with an external instruction-formatted grammar. 
Changes to the SDN infrastructure typically require reconstruction or substantial modification of the grammar and LLM-related control mechanisms.
Second, in large-scale networks with multi-vendor devices, the number of tools and associated function descriptions grows significantly.
This results in lengthy prompts containing extensive tool specifications for each function call or agent invocation, leading to increased token consumption and higher operational costs.
%
%
Finally, the fine-tuning process itself presents additional challenges. Each SDN adaptation requires a dedicated dataset of user requests paired with appropriate SDN tool calls. Consequently, whenever new devices or vendors are added, further adaptation and re-training of the LLM are necessary. In addition, fine-tuning often leads to overfitting, as demonstrated in our manuscript by the baseline of local LLM-based code generation. The fine-tuned model performs well when user inputs closely match the distribution of the request–tool training dataset; however, when users paraphrase their requests, the task execution success rate declines significantly. This sensitivity to linguistic variation limits the robustness and practical applicability of such methods.

To address these limitations, {\name} adopts a protocol-centric design that fundamentally separates language reasoning from device execution.
Rather than relying on handcrafted grammars or embedding detailed tool specifications directly into prompts, which become unscalable in evolving multi-vendor infrastructures, {\name} introduces a structured protocol-level interface that standardizes tool invocation independent of natural language phrasing.
This decoupling eliminates the need for continual grammar updates as devices or APIs change.
Moreover, by abstracting execution into a protocol-governed layer instead of fine-tuning on request-tool pairs, {\name} better preserves the native reasoning capabilities of LLMs and enables reliable closed-loop automation across heterogeneous devices.

\section{{\name}: Design and Implementation}
\label{sec:methods}

In this section, we present the design and implementation of {\name} for optical device control built on MCP, comprising 64 MCP tools across eight representative optical devices and supporting both cloud-hosted commercial and locally deployed open-source LLMs.
{\name} provides a structured orchestration layer in which an LLM serves as the language-driven interface that maps natural language user goals to sequences of protocol-compliant MCP tool invocations across heterogeneous optical devices.

\subsection{MCP-based Agentic System Design}

MCP is an open and standardized interoperability protocol designed to enable structured communication between LLM-based applications and external data sources, tools, and services.
It defines a formal client-server architecture in which the MCP client typically resides on the user side, while MCP servers are deployed on the device side, allowing LLM hosts to systematically discover, access, and utilize contextual resources exposed by devices.
Compared to tool-calling methods, the MCP-based approach is more scalable, enabling remote optical device control with one additional MCP layer.
In {\name}, as shown in Fig.~\ref{fig:overview-diagram}(c), a user initiates interaction by issuing a natural language task (e.g., querying the status of an EDFA), which is received by an MCP client embedded within the host application.
The client forwards the task to the LLM, which interprets the user intent using domain knowledge and selects the relevant and appropriate MCP server(s) based on its published server description, where the MCP server(s) can be located in different locations or sub-networks from the host.
Subsequently, the client retrieves the available tool descriptions and function definitions from the selected MCP server(s) and supplies them to the LLM, which determines the most suitable tool by evaluating semantic similarity between the user task and tool metadata.
The selected tool is then executed by the MCP server, invoking device-specific APIs, monitoring task completion, and returning the execution results to the MCP client. 
Finally, the MCP client relays the results back to the LLM, which processes the output and generates a human-readable natural language response for the user.
Note that the LLM interacting with the MCP client can be either a commercial online model (e.g., GPT-5 or Sonnet~4.5) or a locally hosted open-source model (e.g., Gemma-4-12B or Llama-3.1-8B) with performance, cost, latency, and privacy trade-offs.

This MCP-based design offers several advantages for agentic AI-based optical device control compared with existing approaches~\cite{liu2025first,sun2025automation}.
First, each MCP tool encapsulates a well-defined operational capability, allowing the MCP client to invoke device functions without granting the LLM direct access to low-level device systems or requiring it to generate control code, thereby improving robustness and operational safety.
Rollback and recovery, however, remain tool- and device-dependent, since not all optical control actions are universally reversible. For example, a temporary OSA sweep-parameter change can be rolled back by restoring the previous wavelength range and resolution settings, whereas a ROADM wavelength-selective switch (WSS) reconfiguration may require a device- and topology-specific restoration procedure to avoid disrupting active channels. In practice, access control can be enforced at the MCP server through deployment and tool authorization policies, since device commands are executed through server-side tools rather than through direct LLM access to the device. 
In addition, safety constraints can be embedded directly within the MCP tool layer through device-specific input validation and execution guards. In practical deployments, write operations may reject out-of-range parameter values and unsupported port or channel assignments before any device-side command is issued. This allows operational limits to be enforced according to the constraints of each device and tool, reducing the likelihood of unsafe configurations during autonomous execution. 
Second, the decoupled implementation of MCP clients and servers enables remote device operation across network boundaries, providing increased flexibility in system deployment and access control, compared to tool-calling methods.
For example, a user can run the MCP client on a separate host while keeping the MCP server close to the optical devices, so multi-device actions (e.g., configuring ROADM channels and then measuring spectrum from OSA) are executed locally, and only the structured results are returned to the client.
Finally, MCP provides a uniform communication interface that abstracts vendor-specific protocols, enabling administrators and users to operate systems composed of diverse optical hardware through the same tool interface, even when the underlying devices use different control protocols, without necessitating retraining or fine-tuning of the underlying language models. To add a new optical device to the {\name} system, an MCP server is registered with all relevant optical device control MCP tools, ensuring easy scalability and modularity.
Scalability in {\name} arises from modular server-side tool exposure rather than embedding all device APIs or manuals into a single prompt. In a prompt-driven code-generation workflow, the context can grow substantially as more devices, functions, and vendor-specific documentation are introduced. In contrast, {\name} organizes tools by MCP server, so onboarding a new optical device only requires implementing and registering a new server with its validated tool schemas, while preserving the same client-side orchestration logic. This modular design makes it easier to add and manage devices, and avoids placing all device APIs, manuals, and control logic into a single monolithic prompt for every interaction.
We note that this scalability is architectural rather than a demonstrated deployment at hundreds of devices. Integrating a new optical device still requires device-specific backend development, including implementing the device-specific backend interface and defining the corresponding MCP tool schemas. Once this backend is implemented, however, the LLM-facing interface remains modular: the new device can be exposed as an additional MCP server without retraining the LLM or embedding the full device manual into a monolithic prompt.

\begin{table*}[!t]
\centering
\renewcommand{\arraystretch}{1.1}
\begin{tabular}{|>{\raggedright\arraybackslash}m{5.0cm}|>{\centering\arraybackslash}m{1.3cm}|>{\raggedright\arraybackslash}m{9.0cm}|}
\hline
\textbf{Optical Device} & \textbf{\# of Tools} & \textbf{Example Tools} \\
\hline
Lumentum ROADM & 10 & Set/get EDFA gain; set/get WSS connections/attenuation, ... \\
\hline
Lumentum 400\thinspace{GbE} CFP2 digital coherent optics (CFP2-DCO) & 6 & Set center frequency/output power/operation mode; get config, ... \\
\hline
Optilab LT-12-E-M ARoF Tx & 6 & Set bias voltage/current; get status, ... \\
\hline
APEX Technologies OSA & 26 & Get power/spectrum; set/get measurement parameters, ... \\
\hline
Calient S320 Optical Circuit Switch & 4 & Get port; add/delete connection; delete all connections \\
\hline
DiCon Microelectromechanical Systems (MEMS) 32$\times$32 Optical Switch & 2 & Get connections; set connection \\
\hline
Luna POD2000 Polarimeter & 7 & Set configuration; read polarization, read power, ... \\
\hline
Luna PCD-M02 Polarization Controller & 3 & Reset DAC code; set DAC code; set voltage \\
\hline
\end{tabular}
\normalsize
\caption{List of optical devices and validated MCP tools supported by {\name}.}
\label{tab:mcp-tool-list}
\vspace{-3.0mm}
\end{table*}


\subsection{Implementation of MCP Servers and Tools}

We implement eight MCP servers for eight representative optical devices, as summarized in Table~\ref{tab:mcp-tool-list}:
\textit{(i)} Lumentum ROADM including booster/preamp erbium-doped fiber amplifiers (EDFAs) and 1$\times$20 multiplexer (MUX)/demultiplexer (DEMUX) wavelength-selective switch (WSSs);
\textit{(ii)} Lumentum 400\thinspace{GbE} CFP2-DCO tunable transceiver (TRX);
\textit{(iii)} Optilab analog radio-over-fiber (ARoF) transmitter (Tx) based on an electro-absorption modulator (EAM);
\textit{(iv)} APEX Technologies optical spectrum analyzer (OSA);
\textit{(v)} Calient S320 optical circuit switch (OCS);
\textit{(vi)} DiCon MEMS 32$\times$32 optical switching system;
\textit{(vii)} Luna POD2000 polarimeter; and
\textit{(viii)} Luna PCD-M02 miniature multi-channel piezoelectric actuator driver card with integrated PolaRITE III polarization controller.
For each device, we implement its respective MCP server that exposes a set of tools supporting core operations, including device setup, parameter control, status monitoring, and connection reconfiguration. Note that one tool list above can include many smaller tools, and we implement our tools based on our needs. 
The design architecture of {\name} is inherently extensible to a broad range of optical devices and instruments,
enabling the construction of scalable systems composed of heterogeneous components.

We implement {\name} interacting with two types of LLMs serving for optical device control: commercial online LLMs and locally deployed open-source LLMs:
\begin{itemize}[topsep=3pt, itemsep=3pt,leftmargin=*]
\item
\textbf{MCP with online LLM ({\name}-Online)}, where the MCP client sends tasks to online LLMs.
We consider online models from different platforms\footnote{Model pricing is based on publicly available rates as of February 2026.}, including GPT-4o~mini, GPT-5, Claude Haiku~3.5, and Claude Sonnet~4.5.
GPT-4o~mini has the lowest input cost (\$0.15/\$0.6 per 1M input/output tokens), whereas Haiku~3.5 and Sonnet~4.5 are more expensive (\$1/\$5 and \$3/\$15 per 1M input/output tokens, respectively) due to their larger model capacity.
GPT-5 occupies an intermediate position with a cost of \$1.25/\$10 per 1M input/output tokens, balancing computational cost and reasoning performance.
\item
\textbf{MCP with local LLMs ({\name}-Local)}, which are hosted on a dedicated GPU server accessible on the laboratory network.
We evaluate Gemma-4-12B, Gemma-4-E4B, Gemma-4-E2B, Llama-3.1-8B, and Llama-3.2-3B through an OpenAI-compatible inference endpoint.
None of the locally hosted models is quantized; each is served from its original, unquantized checkpoint using the vLLM inference framework.
Unlike commercial online models, locally deployed LLMs incur no per-token usage costs; instead, expenses are dominated by electricity consumption and one-time investments in GPU and server hardware.
Consequently, the effective token cost of local LLM inference is considered negligible and approximated as zero in our analysis.
\end{itemize}

\section{EXPERIMENTAL EVALUATION}
\label{sec:evaluation}

We compare two workflows for operating real optical devices from natural language task descriptions: \textit{(i)} {\name}, and \textit{(ii)} an LLM-based code generation approach (CodeGen), both within a carefully crafted evaluation benchmark. 

\begin{figure}[!t]
    \centering
    \includegraphics[width=0.98\columnwidth]{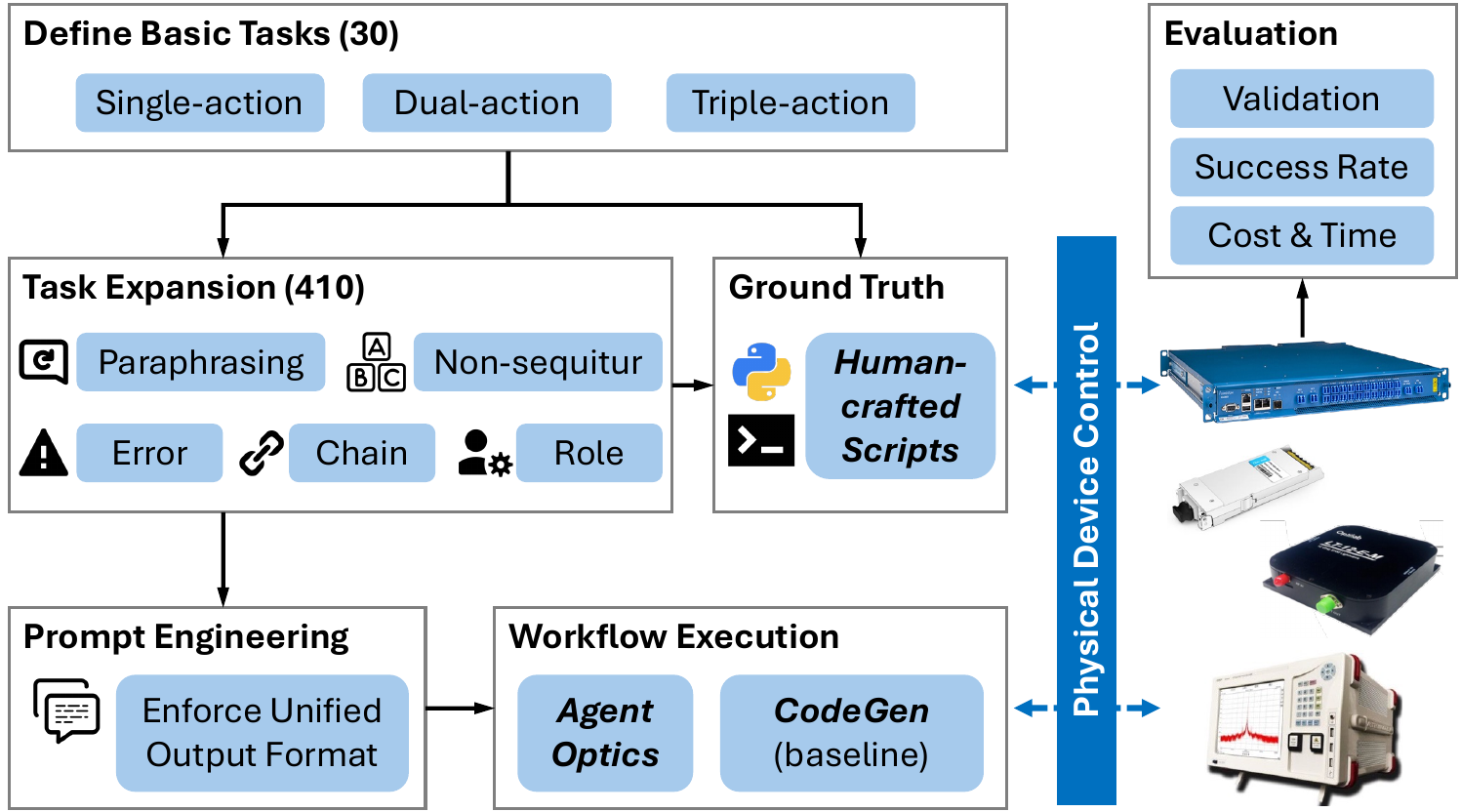}
    \caption{Benchmark workflow for evaluating the performance of {\name} and the CodeGen baseline, where reference ground truth is established using human-crafted scripts that are manually validated on physical devices for correctness.}
    \label{fig:optmcp-eval}
    \vspace{-3.0mm}
\end{figure}

\subsection{Evaluation Benchmark}
\label{sec:evaluation_benchmark}
As shown in Fig.~\ref{fig:optmcp-eval}, the evaluation benchmark begins with basic operational tasks, categorized into single-, dual-, and triple-action tasks.
Each task corresponds to one or more optical device action requests, such as reading device status or configuring operational parameters, expressed in human-readable language.
These tasks are then translated into structured action sequences for execution and evaluation, where prompt engineering is also applied to enforce a unified output format, enabling consistent downstream processing.
We implement device control scripts based on manufacturer manuals and existing codebases to obtain the \textbf{ground truth} device status and operation results (see Fig.~\ref{fig:overview-diagram}(a)), which provides the most accurate and reliable reference for validating optical device operations, and has been extensively used and validated across a series of works~\cite{wang2023dataset,wang2024multi,wang2025scalable,wang2025toward,wang2025scalable_jocn,nishizawa2024semi,xie2025wdm}.
For example, the ROADM control scripts have been used for collecting EDFA gain profile datasets~\cite{wang2023dataset}, as well as modeling and optimizing multi-span link QoT~\cite{raj2024multi,wang2024multi,wang2022multipsan}.
Similarly, the ARoF and OSA control scripts have been used to study the coexistence of fiber sensing, 400\thinspace{GbE}, and 5G signals in field trials~\cite{wang2023field}.

To evaluate the performance of {\name}, each benchmarking task is executed on real optical devices and compared with the ground truth.
Our goal is to compare two practical workflows for  natural-language-driven optical device control under the same hardware and execution environment: structured MCP-based tool invocation and LLM-based code generation.
All workflows run within an identical execution environment, ensuring controlled and reproducible comparisons.
We consider \textbf{\textit{task success rate}} as the primary metric, where the task is labeled as successful when the hardware behavior results based on the task execution match the expected device behavior results. 
More specifically, this includes: 
(\textit{i}) correct tools and invocation order;
(\textit{ii}) correct arguments used for tool invocation; and
(\textit{iii}) execution results meet task intents.
We categorize failure types by analyzing execution result logs and identifying the root causes of failed tasks. Task success itself is determined by whether the observed hardware behavior and returned results match the expected ground-truth execution outcome. We manually inspected all failed executions and additionally spot-checked successful executions using execution traces and device-side results. Although the benchmark contains 410 expanded tasks spanning multiple task complexities and linguistic variants, each task instance is executed once per model in our real-hardware evaluation. Accordingly, the reported results should be interpreted as empirical success rates over a fixed benchmark rather than repeated-trial estimates, and we therefore do not report confidence intervals or variance in this work.

We also consider the \textbf{\textit{average cost}} per task by multiplying the token consumption of each task execution by the corresponding per-token prices of different online LLMs. 
For locally deployed open-source LLMs, the per-token API usage cost is effectively negligible, while the total cost of ownership is instead dominated by electricity consumption and one-time investments in server and GPU hardware. Therefore, the per-task cost for local models is approximated as zero only in the API-usage sense in our evaluation.
Moreover, we record the \textbf{\textit{execution time}} per task, including the time spent on task input, tool selection, device control, and retrieval and analysis of the returned value.
For tasks involving multiple actions, the execution time is measured as the total time required to complete all actions.

\subsection{Agentic Optical Device Control Tasks}
\label{sec:evaluation_tasks}
To systematically evaluate the performance of {\name}, we develop a benchmark consisting of a comprehensive set of carefully designed user tasks, as shown in Fig.~\ref{fig:optmcp-eval}.
The benchmark evaluates agent performance across multiple dimensions, including the number of device-control APIs required to execute a user command, robustness to diverse natural language inputs, error-handling capability, and support for multi-vendor device operation and coordination.
This benchmark spans a representative set of optical hardware, including a Lumentum ROADM, a 400\thinspace{GbE} coherent CFP2-DCO, an OptiLab ARoF transmitter, and an APEX OSA, which are commonly used together across a wider range of optical networking and experimentation workflows.
The benchmark includes user tasks that involve one to three actions (e.g., invoking vendor-specific APIs) spanning multiple devices.
For example, a single-action task may read the signal spectrum from an OSA, while a dual-action task may combine two related or unrelated operations within a single prompt, such as reading signal spectrum from an OSA followed by setting the output power of an EDFA.

We further extend each task to multiple representative variants, designed to evaluate the robustness of the agentic AI framework under conditions commonly encountered in day-to-day optical device and network control.
Specifically, we consider five types of \textit{task variants} (Table~\ref{tab:benchmark}):
\begin{itemize}[topsep=3pt, itemsep=3pt, leftmargin=*]
    \item
    \textbf{\textit{Paraphrasing}} evaluates semantic understanding by testing whether the agent can recognize differently worded instructions with identical intent (e.g., ``\textit{OSA measure}'' $\Leftrightarrow$ ``\textit{OSA data recording}'');
    \item
    \textbf{\textit{Non-Sequitur}} assesses robustness to irrelevant or incoherent instructions that should not result in valid actions (e.g., ``\textit{OSA measure, watch TV}'');
    \item
    \textbf{\textit{Error}} tests the agent's ability to detect and reject invalid or unsafe commands (e.g., ``\textit{Set OSA wavelength to 0}'');
    \item
    \textbf{\textit{Chain}} measures multi-step reasoning and state consistency across sequentially dependent commands (e.g., ``\textit{Set EDFA gain, then read EDFA gain}'');
    \item
    \textbf{\textit{Role}} evaluates contextual and role-based instruction following, where task execution depends on an assigned role (e.g., ``\textit{Act as the service provider, OSA measure}'').
\end{itemize}

The benchmark begins with 30 basic task templates, consisting of 10 single-action, 10 dual-action, and 10 triple-action tasks. These templates are systematically expanded into a total of 410 tasks spanning multiple difficulty levels and interaction patterns.
Specifically, for each basic \textit{single-action} task, we generate five paraphrasing variants and five non-sequitur variants, as well as three error variants and three role-conditioned variants, yielding $10\times\left(5+5+3+3\right)=160$ expanded tasks. 
The error and role-conditioned variants are applied only to single-action tasks to control the evaluation variables. This design isolates the effect of invalid/missing inputs or role-conditioned prompts, whereas applying these variants to multi-action tasks would confound error handling with multi-step coordination.
For each basic \textit{dual-action} task, we generate five paraphrasing variants and five non-sequitur variants, as well as 10 chained variants constructed by converting the two-step format into a single sequential instruction that includes both steps (e.g., ``\textit{(1) set ...; (2) get/read ...}'' $\Rightarrow$ ``\textit{set ... and then get/read ...}''), each with five paraphrasing variants, yielding $10\times\left(5+5\right)+10\times5=150$ expanded tasks. 
For each basic \textit{triple-action} task, we generate five paraphrasing variants and five non-sequitur variants, yielding $10\times\left(5+5\right)=100$ expanded tasks. 
In total, this yields 410 expanded tasks that are used for evaluating the performance of {\name}.

We emphasize that the 410-task benchmark is not intended to represent 410 fully independent control objectives. Rather, it expands 30 basic tasks into varied linguistic and interaction settings to assess the robustness of {\name} under realistic user-wide variations. Accordingly, the expanded benchmark should be interpreted as a robustness-oriented evaluation built upon 30 basic task templates spanning single-, dual-, and triple-action optical-device control. The basic tasks capture the diversity of underlying control objectives, while the derived variants stress the stability and reliability of the agent under realistic usage conditions. 


\begin{table*}[ht]
\centering
\renewcommand{\arraystretch}{1.2}

\begin{tabular}{|
  >{\arraybackslash}m{1.5cm} |
  >{\arraybackslash}m{3.0cm} |
  >{\arraybackslash}m{11.5cm} |
}
\hline
\textbf{Type} & \textbf{Description} & \textbf{Task example} \\
\hline
Paraphrasing &
Same meaning, different phrases &
\begin{itemize}[leftmargin=*, itemsep=0pt, topsep=0pt, parsep=0pt, partopsep=0pt]
    \item Operate the CFP2 so that port cfp2-opt-1-1 has an output target power setting of $-$5\thinspace{dBm}.
    \item Using the CFP2, adjust the output target power parameter on port cfp2-opt-1-1 to $-$5\thinspace{dBm}.
\end{itemize} \\
\hline
Non-sequitur &
Adding unrelated information to the task &
\begin{itemize}[leftmargin=*, itemsep=0pt, topsep=0pt, parsep=0pt, partopsep=0pt]
    \item Set CFP port cfp2-opt-1-1 power to $-$5\thinspace{dBm}; the bench mat has a curled corner.
\end{itemize} \\
\hline
Error &
Task with wrong or lost value &
\begin{itemize}[leftmargin=*, itemsep=0pt, topsep=0pt, parsep=0pt, partopsep=0pt]
    \item Missing power value: on the CFP2, set output target power on port cfp2-opt-1-1.
    \item Wrong power value: on the CFP2, set output target power on port cfp2-opt-1-1 to $-$100\thinspace{dBm}.
\end{itemize} \\
\hline
Chain &
Sequential related tasks &
\begin{itemize}[leftmargin=*, itemsep=0pt, topsep=0pt, parsep=0pt, partopsep=0pt]
    \item First set CFP2 port cfp2-opt-1-1 output target power to $-$4\thinspace{dBm}, then read CFP2 output power.
\end{itemize} \\
\hline
Roles &
Task tone as service provider or user &
\begin{itemize}[leftmargin=*, itemsep=0pt, topsep=0pt, parsep=0pt, partopsep=0pt]
    \item You are an optical device user; set CFP port cfp2-opt-1-1 power to $-$5\thinspace{dBm}.
\end{itemize} \\

\hline
\end{tabular}
\caption{Five representative task variants evaluated in the agentic optical device control benchmark.}
\label{tab:benchmark}
\vspace{-3.0mm}
\end{table*}

\subsection{Baseline: LLM-based Code Generation (CodeGen)}
\label{sec:baselines}

We evaluate {\name} against LLM code generation workflows that rely on either user manuals provided by the device manufacturer or pre-existing code libraries for direct device operation, as illustrated in Fig.~\ref{fig:overview-diagram}(b). 
The code generated by the LLM is then executed in a clean and isolated runtime environment on actual devices.
The returned output and execution logs/errors from the device are captured for evaluation.
The task success rate for the CodeGen baseline is defined identically to that for {\name} described in Section~\ref{sec:evaluation_benchmark}, where only hardware returned results that meet the expected outcome are considered successful. 
We evaluate both online high-capacity models and a locally hosted model supporting privacy-preserving deployments:
\begin{itemize}[topsep=3pt, itemsep=3pt, leftmargin=*]
\item
\textbf{LLM-based code generation with online LLMs (CodeGen-Online)} uses online LLMs (e.g., Claude Sonnet~4.5) to generate control code in Python based on the device documentation.
The LLM receives task descriptions, device connection information (e.g., IP address, serial ports, USB identifiers), and relevant materials in one of the two forms:
(\textit{i}) vendor-provided manuals, including APIs/code that is part of the manual (CodeGen-Online with manual), or
(\textit{ii}) reference code libraries for each target device (CodeGen-Online with code).
The LLM then produces executable code to control the target optical devices. 
For the reference-code setting, we provide the same MCP server source code that underlies {\name}, which contains the corresponding device-control implementations and validation logic, to enable a fair comparison with the MCP approach at the backend-logic level.
\item
\textbf{LLM-based code generation with local LLMs (CodeGen-Local)} uses CodeLlama-7b-hf~\cite{roziere2024codellama}, which is a 7 billion-parameter foundation model in the Code Llama family developed for general code synthesis and understanding, capable of generating programming code from natural language or partial code prompts.
However, small local models such as CodeLlama-7b-hf usually exhibit significantly weaker code generation performance compared to online models.
To mitigate this limitation, we applied LoRA fine-tuning~\cite{hu2022lora} using a local dataset consisting of user-issued device control commands (e.g., requesting an OSA spectrum measurement) paired with their corresponding control code.
This task-specific adaptation transforms a general-purpose local LLM into a specialized optical device control model, substantially improving control accuracy and execution success rates.
The LoRA fine-tuning employs a learning rate of $3 \times 10^{-4}$ over 100 epochs. 
\end{itemize}

For CodeGen-Local, the LoRA fine-tuning corpus is constructed from 30 basic tasks described in Section~\ref{sec:evaluation_tasks}, while evaluation is performed on the 410 expanded task variants derived from these tasks. As a result, the CodeGen-Local results should be interpreted as measuring robustness and generalization to linguistic and interaction variations of seen root tasks, rather than performance on entirely unseen task families. We adopt this setting to provide a stronger local code-generation baseline for comparison with {\name}.

We note that this comparison is designed as a workflow-level evaluation rather than a schema-identical baseline match. In particular, {\name} invokes pre-defined MCP tools through structured tool schemas, whereas CodeGen must generate executable code that uses the same underlying device-control implementation. By providing the same MCP server source code in the CodeGen-Online with code setting, we substantially reduce asymmetry between the two approaches at the level of backend logic, validation routines, and hardware access.



\section{RESULTS}
\label{sec:results}


We evaluate the performance of {\name} using different commercial online and locally deployed open-source LLMs under the benchmark described in Section~\ref{sec:evaluation}, compare {\name} against CodeGen baselines, analyze agentic AI execution error types, and finally compare the cost and execution time of {\name} with the CodeGen baseline.

\subsection{{\name} with Online and Local LLMs}

\begin{figure}[!t]
    \centering
    \includegraphics[width=0.95\columnwidth]{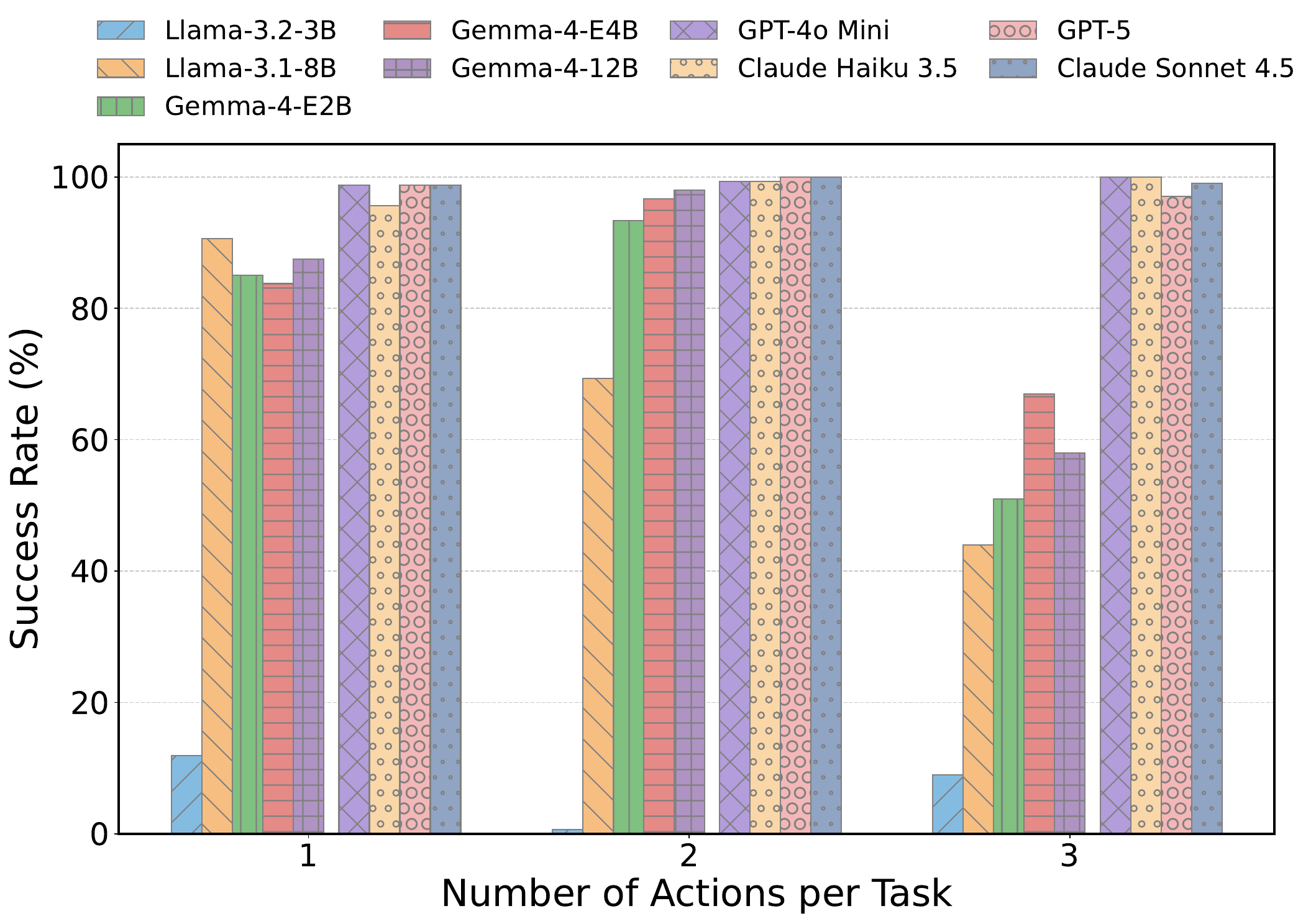}
    \caption{Task success rate achieved by {\name} across varying task complexities using five locally hosted Gemma and Llama models and four online LLMs. The two-action bar aggregates direct and chained two-action tasks.}
    \label{fig:mcp_accuracy_tool_number}
    \vspace{-3.0mm}
\end{figure}

Fig.~\ref{fig:mcp_accuracy_tool_number} shows the success rate achieved by {\name} for tasks with varying numbers of actions under different online ({\name}-Online) and local ({\name}-Local) LLMs.
Overall, commercial online models achieve success rates from {95.6\%} to {98.8\%} for single-action tasks, from {99.3\%} to {100.0\%} for dual-action tasks, and from {97.0\%} to {100.0\%} for triple-action tasks.
The Gemma models achieve {83.8\%} to {87.5\%} for single-action tasks, {93.3\%} to {98.0\%} for the combined dual-action and chain group, and {51.0\%} to {67.0\%} for triple-action tasks. Llama-3.1-8B achieves {90.6\%}, {69.3\%}, and {44.0\%}, respectively, showing a larger degradation as action depth increases. Llama-3.2-3B reaches only {11.9\%}, {0.7\%}, and {9.0\%}; its failures are predominantly missing or unparsable tool calls, rather than device-side execution failures. These results highlight the remaining challenge of reliable multi-tool coordination for smaller local models.
Notably, GPT-4o~mini demonstrates consistently strong performance across all task complexities despite its lower cost than the larger online models.

\begin{figure}[!t]
    \centering
    \includegraphics[width=0.95\columnwidth]{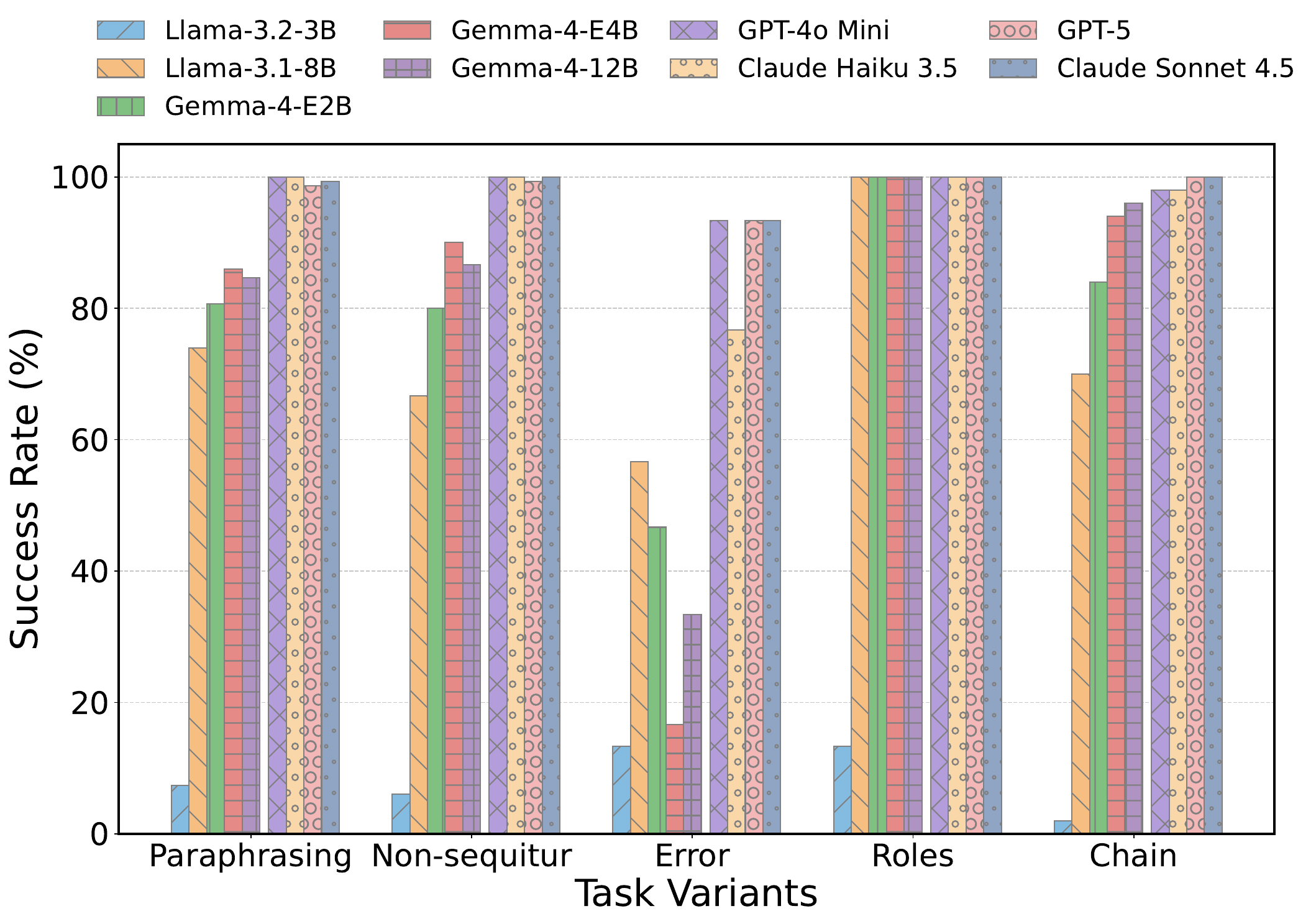}
    \caption{Task success rate achieved by {\name} across different task variants using five locally hosted Gemma and Llama models and four online LLMs.}
    \label{fig:mcp_accuracy_tool_variation}
    \vspace{-3.0mm}
\end{figure}

Fig.~\ref{fig:mcp_accuracy_tool_variation} shows the success rate achieved by {\name} across different task variants described in Section~\ref{sec:evaluation_benchmark} (see also Table~\ref{tab:benchmark}).
The online models are robust to paraphrasing ({98.7\%} to {100.0\%}) and non-sequitur prompts ({99.3\%} to {100.0\%}). The Gemma models achieve {80.7\%} to {86.0\%} and {80.0\%} to {90.0\%}, respectively; Llama-3.1-8B achieves {74.0\%} and {66.7\%}, while Llama-3.2-3B achieves {7.3\%} and {6.0\%}. Role prompts are solved by all Gemma models and Llama-3.1-8B, but only {13.3\%} are solved by Llama-3.2-3B. The error variant remains the most difficult setting for local models: the Gemma models achieve {16.7\%} to {46.7\%}, Llama-3.1-8B achieves {56.7\%}, and Llama-3.2-3B achieves {13.3\%}, compared with {76.7\%} to {93.3\%} for the online models. Although explicit error-handling mechanisms are implemented within the MCP tools integrated in {\name}, missing or incomplete input parameters can still cause an LLM to invoke an incorrect API or tool.

\subsection{{\name} and CodeGen Performance Comparison}

\begin{figure}[!t]
    \centering
    \includegraphics[width=0.95\columnwidth]{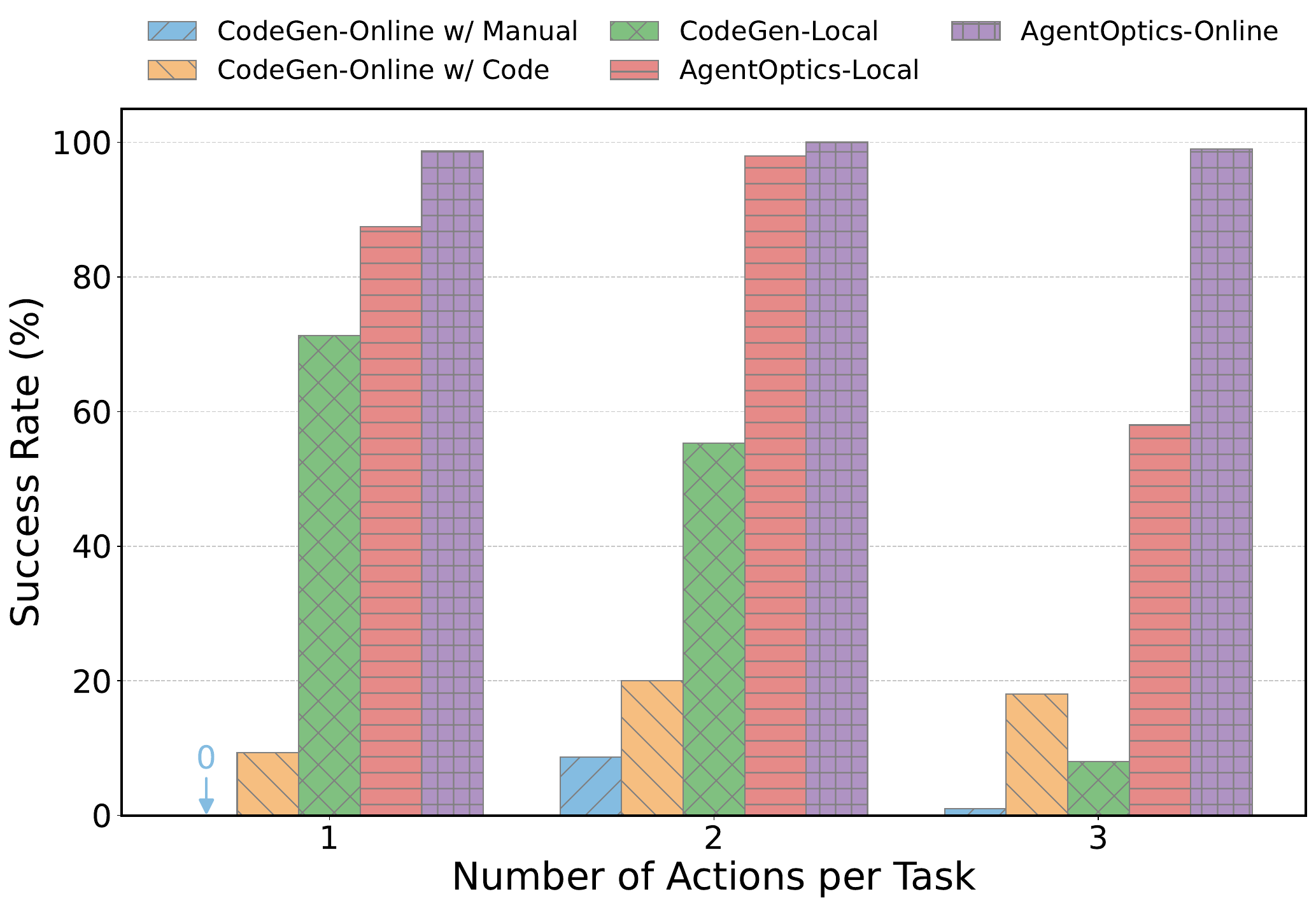}
    \caption{Task success rate achieved by {\name} using AgentOptics-Local (Gemma-4-12B) and AgentOptics-Online (Claude Sonnet~4.5), compared with the CodeGen baseline. The two-action bar aggregates direct and chained two-action tasks.}
    \label{fig:all_accuracy_tool_number}
    \vspace{-3.0mm}
\end{figure}

Next, we compare the performance of {\name} against the CodeGen baseline using both an online model (Claude Sonnet~4.5) and a locally deployed, domain fine-tuned model (CodeLlama-7b-hf, see Section~\ref{sec:baselines}).
Fig.~\ref{fig:all_accuracy_tool_number} shows the corresponding task success rates for single-, dual-, and triple-action tasks. {\name}-Online attains near-perfect performance, achieving {98.8\%} to {100.0\%}. The representative AgentOptics-Local configuration (Gemma-4-12B) achieves {87.5\%} for single-action tasks, {98.0\%} for the combined dual-action and chain group, and {58.0\%} for triple-action tasks.
On the other hand, the CodeGen baseline exhibits substantially lower success rates across all task complexities. Notably, CodeGen-Local shows a clear degradation as the number of required actions increases ({71.3\%} $\to$ {55.3\%} $\to$ {8.0\%}).
For example, on single-action tasks, CodeGen-Local achieves a success rate of {71.3\%}, whereas {\name}-Online maintains a success rate of 98.8\%.
This performance gap is more significant for complex tasks: for triple-action tasks, CodeGen-Local achieves a success rate of only {8.0\%}, whereas {\name}-Online maintains a success rate of {99.0\%}.
Notably, CodeGen-Online with code and manuals performs even worse, achieving success rates of at most {20.0\%} because of the lack of domain-specific knowledge in the online model.
This low success rate does not imply that the online model lacks general coding capability; rather, it reflects the difficulty of synthesizing directly executable device-control code from long manuals or reference code under strict hardware-interface constraints. The generated scripts often contain small but execution-critical errors, such as incorrect function names and invalid object attributes.

In contrast, {\name} exposes the same backend device-control logic through structured MCP tools, so the LLM primarily selects validated tools and arguments instead of reconstructing device-specific code.
Because the CodeGen-Online with code baseline is provided with the same backend device-control implementation used by the MCP servers, this comparison suggests that the observed improvement mainly comes from structured tool invocation and schema-guided parameter binding, rather than from using a different or more capable device-control backend.
In contrast, CodeGen-Local outperforms the CodeGen-Online variants on low-complexity tasks (e.g., {71.3\%} vs. {9.4\%} for single-action tasks), primarily due to the benefits of LoRA fine-tuning.
However, this advantage comes at the expense of scalability, as the local LLM must be fine-tuned every time new devices or features are added. 

\begin{figure}[!t]
    \centering
    \includegraphics[width=0.45\textwidth]{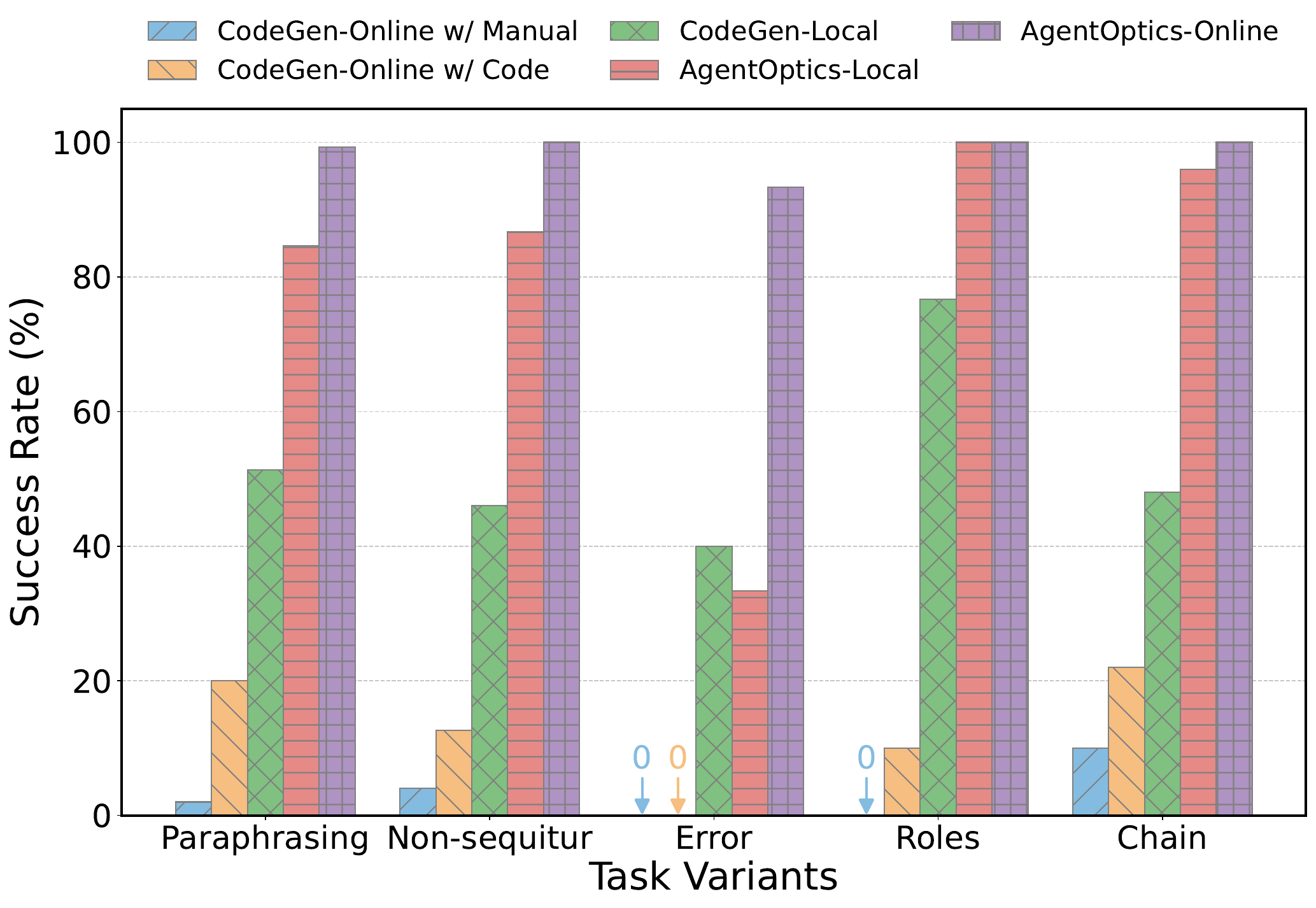}
    \caption{Task success rate achieved by {\name} using AgentOptics-Local (Gemma-4-12B) and AgentOptics-Online (Claude Sonnet~4.5), compared with the CodeGen baseline.}
    \label{fig:all_accuracy_tool_variation}
    \vspace{-3.0mm}
\end{figure}

Fig.~\ref{fig:all_accuracy_tool_variation} reports success rates across paraphrasing, non-sequitur, error, role, and chain variants. {\name}-Online achieves {93.3\%} to {100.0\%} across all variants. The representative AgentOptics-Local configuration (Gemma-4-12B) achieves {84.7\%} on paraphrasing, {86.7\%} on non-sequitur prompts, {33.3\%} on error tasks, {100.0\%} on role prompts, and {96.0\%} on chain tasks.
The CodeGen baseline is substantially weaker on the error and chain variants: both CodeGen-Online variants achieve {0.0\%} on error tasks, and CodeGen-Online with code reaches only {22.0\%} on chain tasks. CodeGen-Local benefits from LoRA fine-tuning, reaching {51.3\%} on paraphrasing tasks compared with {20.0\%} for CodeGen-Online with code, but it remains below {\name}-Online across the evaluated variants.

\subsection{Cost Efficiency and Execution Time}

\begin{figure}[!t]
    \centering
    \includegraphics[width=0.99\columnwidth]{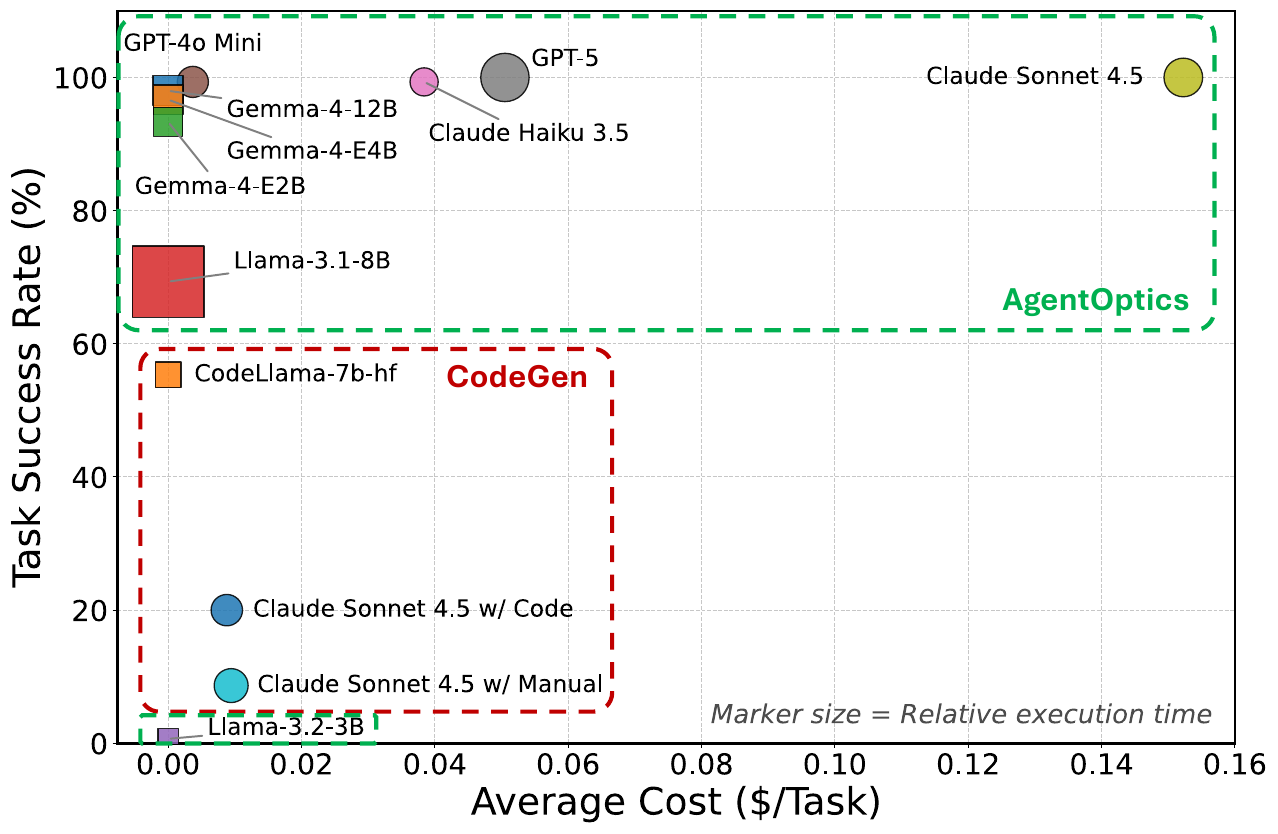}
    \caption{Trade-off between success rate and average cost for the combined direct dual-action and chained-task subset with {\name} and the CodeGen baseline using locally hosted (squares) and online (circles) LLMs.
    Marker size indicates the relative average execution time.}
    \label{fig:mcp_cost_vs_tools}
    \vspace{-3.0mm}
\end{figure}

Fig.~\ref{fig:mcp_cost_vs_tools} shows the trade-offs between task success rate and average cost (\$/task) achieved by {\name} and the CodeGen baseline on the combined direct dual-action and chained-task subset with different LLMs.
Commercial online LLMs show a range of (\textit{cost}, \textit{success rate}) trade-offs: GPT-5 (\$0.051/task, 100.0\%) and Claude Sonnet~4.5 (\$0.152/task, 100.0\%) achieve the highest success rates on this subset at higher cost, whereas GPT-4o~mini (\$0.004/task, 99.3\%) achieves nearly the same performance at substantially lower cost.
The locally hosted Gemma models achieve {93.3\%} to {98.0\%} success on the combined dual-action and chain subset at near-zero API cost. Llama-3.1-8B reaches {69.3\%}, whereas Llama-3.2-3B reaches {0.7\%}; all five local deployments use unquantized model checkpoints, so these comparisons do not conflate reduced numerical precision with model capability. The total cost of ownership is instead dominated by local hardware and energy use.

In contrast, the CodeGen baseline exhibits both limited accuracy (e.g., only up to 55\%) and cost efficiency.
Notably, {\name} is not cost-efficient when paired with high-cost models such as Claude Sonnet~4.5, even compared to both CodeGen-Online variants with code and manuals (\$0.010/task), which require extensive device manuals, often tens to hundreds of pages, or reference codebases consisting of hundreds to thousands of lines of code.
For example, {\name} using Claude Sonnet~4.5 and GPT-5 reaches 100.0\% success on this subset, whereas GPT-4o~mini reaches 99.3\% at a 41$\times$ lower cost than Claude Sonnet~4.5. This suggests that for these structured tool-based control tasks, substantially higher cost may provide only a small incremental benefit.
We also report the relative execution time of different methods in Fig.~\ref{fig:mcp_cost_vs_tools}, indicated by the size of the circles.
On average, for {\name}, execution time varies significantly across models. 
GPT-5 ({19.4}\thinspace{sec}) and Claude Sonnet~4.5 ({12.4}\thinspace{sec}) show longer runtimes on this subset than GPT-4o~mini ({8.0}\thinspace{sec}) and the Gemma models ({6.9}\thinspace{sec} to {7.5}\thinspace{sec}). Llama-3.1-8B averages {42.6}\thinspace{sec}, reflecting repeated agent-loop attempts on difficult tool calls, while Llama-3.2-3B averages {3.4}\thinspace{sec} because it often terminates without an executable call.
When comparing {\name} with the CodeGen baseline using Claude Sonnet~4.5, {\name} incurs additional latency ({12.4}\thinspace{sec} versus {8.2}\thinspace{sec} for CodeGen-Online-Code) due to potentially multiple rounds of communication between the client and LLM running the MCP-based method.

\begin{table*}[!t]
\centering
\begin{tabular}{p{1.5cm} p{3.3cm} p{1.2cm} p{10cm}}
\hline
\textbf{Methods} & \textbf{Failure Category} & \textbf{Percentage} & \textbf{Example Description} \\
\hline
CodeGen & Syntax error & 20.36\% &
Contains invalid syntax (e.g., \texttt{IP=''192.168.0.2}). \\

CodeGen & Call non-existing function & 15.08\% &
Calls an invalid function (e.g., calls \texttt{AP2XXX.get\_powower}, which does not exist). \\

CodeGen & Import non-existing library & 11.18\% &
Imports an undefined library (e.g., \texttt{import lab\_api}, which is undefined). \\
\hline
{\name} & Wrong tool & 89.09\% &
Calls the wrong tool (e.g., \texttt{arof\_get\_power} instead of \texttt{arof\_read\_power}). \\

{\name} & Missing tool & 10.91\% &
Required tools are not invoked (expected OSA-related tools, but called none). \\


\hline
\end{tabular}
\caption{Reasons and examples for CodeGen and MCP-based {\name} execution failures. The percentages are calculated with respect to the total number of failures within each approach.}
\label{tab:error_type}
\vspace{-3.0mm}
\end{table*}

\subsection{Agentic AI Execution Error Types}

{\name} handles input errors naturally through proper error-handling design. For example, within MCP tool implementations, validating input parameters is essential to detect missing or incorrect values. In such cases, the {\name} system can either apply default parameter values or prompt the user to provide correct inputs.
To understand the remaining failures under this benchmark, Table~\ref{tab:error_type} summarizes common failure modes observed when applying {\name} and the CodeGen baseline.
For CodeGen, failures primarily stem from software-level issues, including importing non-existent libraries, invoking invalid functions or class attributes, and generating syntactically incorrect code.
For example, the prompt with ``\textit{set the target output power of a CFP2 module to $-${5}\thinspace{dBm}}'' leads to the error ``\textit{ERROR: \texttt{CFP2} object has no attribute \texttt{set\_output\_power}}''.
Such errors indicate limitations in CodeGen's ability to accurately reason about programming language syntax and library semantics under complex task requirements.
In contrast, {\name} failures are mainly related to tool orchestration, such as missing required tool invocations, incorrect tool naming due to formatting inconsistencies, calling undefined MCP tools, or executing only a subset of the tools necessary to complete a task.
For example, the prompt ``\textit{Set the OSA start wavelength to 1545 nm}'' requires calling the tool \texttt{osa\_set\_start\_wavelength}, but {\name} instead invokes the tool \texttt{osa\_set\_center\_wavelength}.
These failure modes suggest that while {\name} can often reason correctly about individual tools, it still struggles with reliably coordinating multiple tools.



\section{CASE STUDIES}
\label{sec:casestudies}

{\name} is not limited to high-fidelity control of individual optical devices under diverse natural-language inputs, but also extends to more comprehensive network production scenarios, including optical link configuration, autonomous channel QoT optimization, quantum polarization control, and automatic fiber cut event detection. We first show an LLM-assisted MCP tool generation from legacy device-control code, and then demonstrate five representative runtime orchestration case studies.
In the first case study, the LLM is used to synthesize MCP-compatible tool interfaces from device-control code to support rapid device onboarding. Across the remaining five runtime case studies, the LLM is not used only as a natural-language front end to trigger fixed commands, but also performs task-level orchestration over validated MCP tools, including interpreting operator intent, selecting relevant device tools, ordering actions across devices, reasoning about cross-device dependencies, and using returned measurements to determine subsequent steps when feedback is required.
We highlight the specific reasoning role of the LLM at the beginning of each case study below.

\subsection{LLM-Assisted MCP Tool Generation for Rapid Device Onboarding}

We first demonstrate the extensibility of {\name} through an LLM-assisted workflow for generating MCP tools from legacy device-control code. Unlike the remaining case studies, which focus on runtime orchestration over pre-built tools, this case study examines whether an LLM can help reduce the manual engineering effort required to onboard additional device capabilities into the {\name} framework. Because our available hardware is limited, we use the APEX OSA as a representative example and frame this study as an onboarding-workflow demonstration rather than a benchmark on previously unseen hardware classes.

\begin{figure}[!t]
    \centering
    \includegraphics[width=0.95\columnwidth]{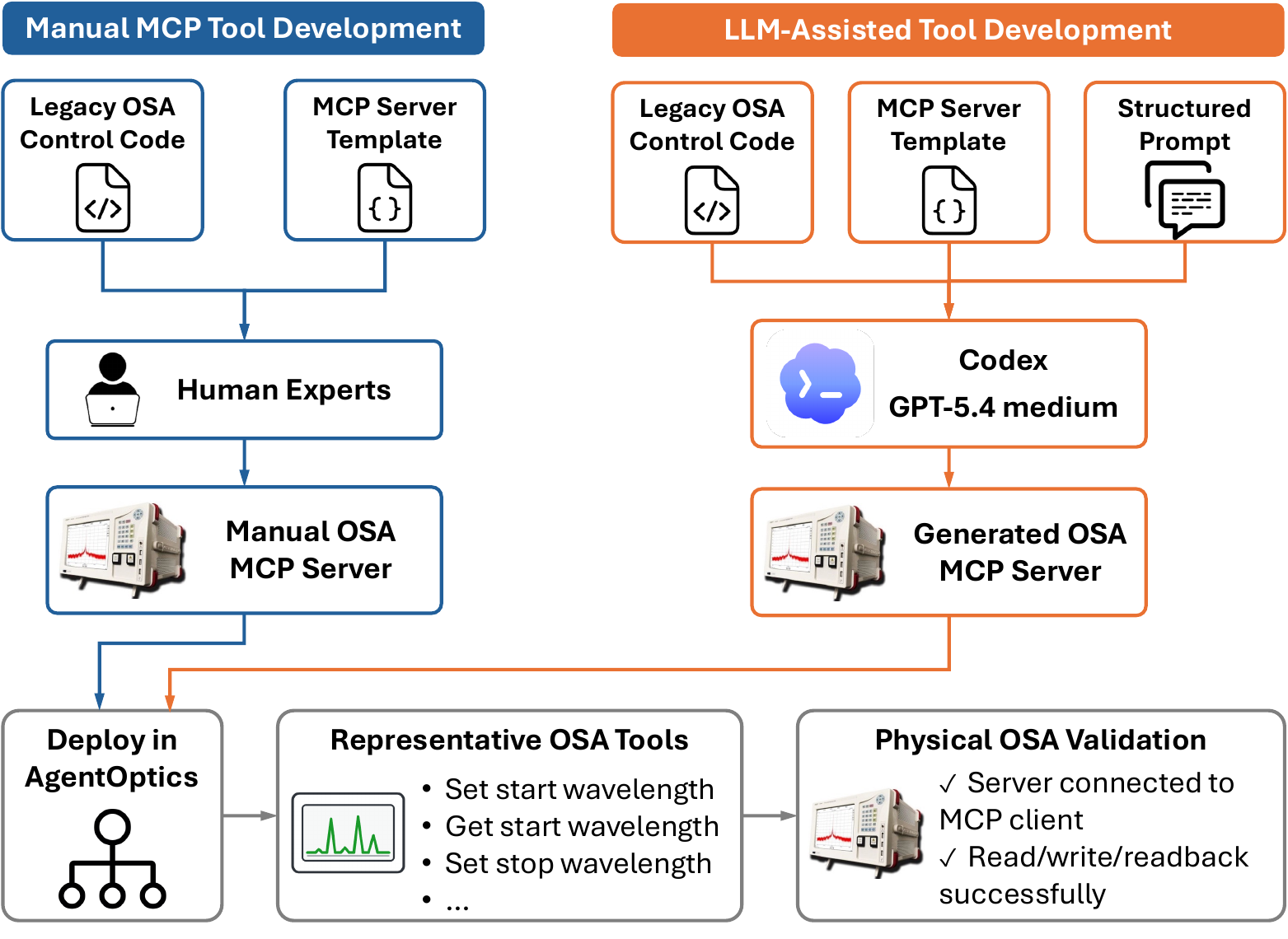}
    \caption{Representative OSA MCP tool onboarding workflows for manual and LLM-assisted MCP server generation.}
    \label{fig:diagram_demo0}
    \vspace{-3.0mm}
\end{figure}

Specifically, we provide GPT-5.4 medium in Codex with the original OSA control source code, an existing MCP server implementation for another device as a template, and a structured prompt describing the required MCP format and validation behavior. The generation task is restricted to a small set of wavelength-control functions that are commonly used in optical experiments. The model then produces a new MCP server that wraps the corresponding OSA backend functions while preserving the server structure required by {\name}. Device-specific access information, such as IP address or credentials, is intentionally left as placeholders and filled in only during local deployment.

As illustrated in Fig.~\ref{fig:diagram_demo0}, we compare this LLM-assisted onboarding workflow with the conventional manual development path for the representative OSA example. We validate the generated server on the physical OSA through the same MCP client interface used in the rest of the paper. Representative read and write queries confirm that the generated server can connect to the MCP client, execute wavelength configuration and readback operations on the instrument, and return structured responses in the format expected by {\name}. This result shows that {\name} can be extended not only through manual MCP server development, but also through an LLM-assisted tool-synthesis workflow that translates legacy device-control code into MCP-compatible interfaces with limited manual effort. Although this experiment does not establish universal generalization to arbitrary unseen hardware, it demonstrates a practical path for onboarding additional optical device capabilities into {\name}.

\subsection{DWDM Link Setup and Performance Monitoring}

\begin{figure}[!t]
    \centering
    \includegraphics[width=0.95\columnwidth]{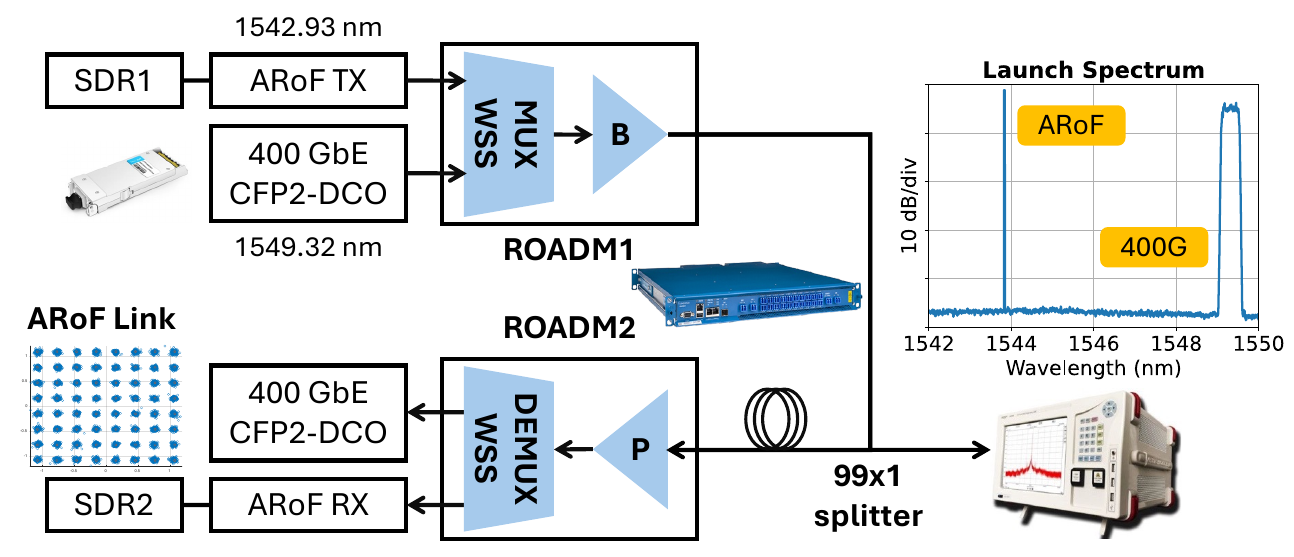}
    \caption{Diagram for DWDM link configuration with ARoF and 400\thinspace{GbE} signals.}
    \label{fig:diagram_demo1}
    \vspace{-3.0mm}
\end{figure}

To evaluate {\name}'s capability to coordinate multi-vendor multi-device optical network operation, we establish a DWDM link that integrates Lumentum ROADMs with coherent 400\thinspace{GbE} and ARoF subsystems.
This configuration, shown in Fig.~\ref{fig:diagram_demo1}, demonstrates end-to-end wavelength provisioning with coordinated control across multiple signal types. 
The topology consists of two ROADM units (ROADM1 and ROADM2) interconnected through a 99$\times$1 optical splitter, where the 1\% port is directed to an OSA for real-time monitoring while 99\% of the optical power propagates through a {20}\thinspace{km} fiber spool connected to ROADM2.
Each ROADM unit includes both MUX/DEMUX WSS modules as well as booster (B) and pre-amplifier (P) EDFAs.
In this experiment, we configure only the required subset of the components, using the MUX WSS and booster EDFA on ROADM1 and the DEMUX WSS and pre-amplifier EDFA on ROADM2. 
This architecture enables wavelength routing with optical amplification for span loss compensation. 

Using an agentic workflow, {\name} executes a coordinated multi-device control sequence from an operator intent to provision the two optical channels described below.
Although the per-channel parameters are specified in the user intent, {\name} must autonomously reason about the cross-device dependencies, selecting and ordering the appropriate MCP tools across heterogeneous vendors, mapping each spectral assignment to the correct WSS port on both the MUX and DEMUX sides, aligning the CFP2-DCO center frequency with the WSS passband, and deciding which monitoring tools to invoke. This multi-vendor, multi-device orchestration is not reducible to a fixed script, since the same operator intent on a different topology would yield a different tool-invocation plan.
In this case study, the agent interacts with four devices (ROADMs, OSA, CFP2-DCO, and ARoF Tx) and invokes nine tools.
Note that although {\name} provides a spectrum measurement tool, it is not invoked here: high-resolution spectrum acquisition would require a large numerical array that significantly increases token consumption in the LLM context.
Instead, the OSA tools only configure the sweep window, and spectrum acquisition is performed outside the {\name} workflow. 
Specifically, it configures an ARoF path over {193.9 to 194.45}\thinspace{THz} on port~1 and a CFP2 path over {193.4 to 193.7}\thinspace{THz} on port~20 on both ROADMs, sets OSA sweep to {1540 to 1550}\thinspace{nm}, sets the CFP2 center frequency to {193.5}\thinspace{THz}, and applies {99}\thinspace{mA} current and $-${0.9}\thinspace{V} bias to the ARoF Tx.

With these WSS paths and endpoint settings, the link carries two optical channels: an ARoF channel operating at {1542.92}\thinspace{nm} ({194.3}\thinspace{THz}) for analog radio signal transmission and 400\thinspace{GbE} at {1549.32}\thinspace{nm} ({193.5}\thinspace{THz}) by the CFP2-DCO for high-speed coherent data transmission.
Both channels are routed through the ROADM WSS modules with appropriate power levels and port assignments.
Performance monitoring successfully confirms provisioning with a measured OSNR of {32.6}\thinspace{dB} for the 400\thinspace{GbE}  signal by the CFP2-DCO and an EVM of 3.91\% for the ARoF signal, while the launch spectrum shows both channels at their designated wavelengths in Fig.~\ref{fig:diagram_demo1}.
In this DWDM setting, potential optical crosstalk can be managed by {\name} through coordinated WSS passband configuration, channel attenuation control, and OSA-based spectrum monitoring; more advanced closed-loop crosstalk mitigation, such as automatic passband adjustment or multi-channel power rebalancing, can be supported by adding corresponding MCP tools such as GNPy or ML-based digital twin.

\subsection{Wideband 5G ARoF Link Characterization}

\begin{figure}[!t]
    \centering
    \includegraphics[width=0.98\columnwidth]{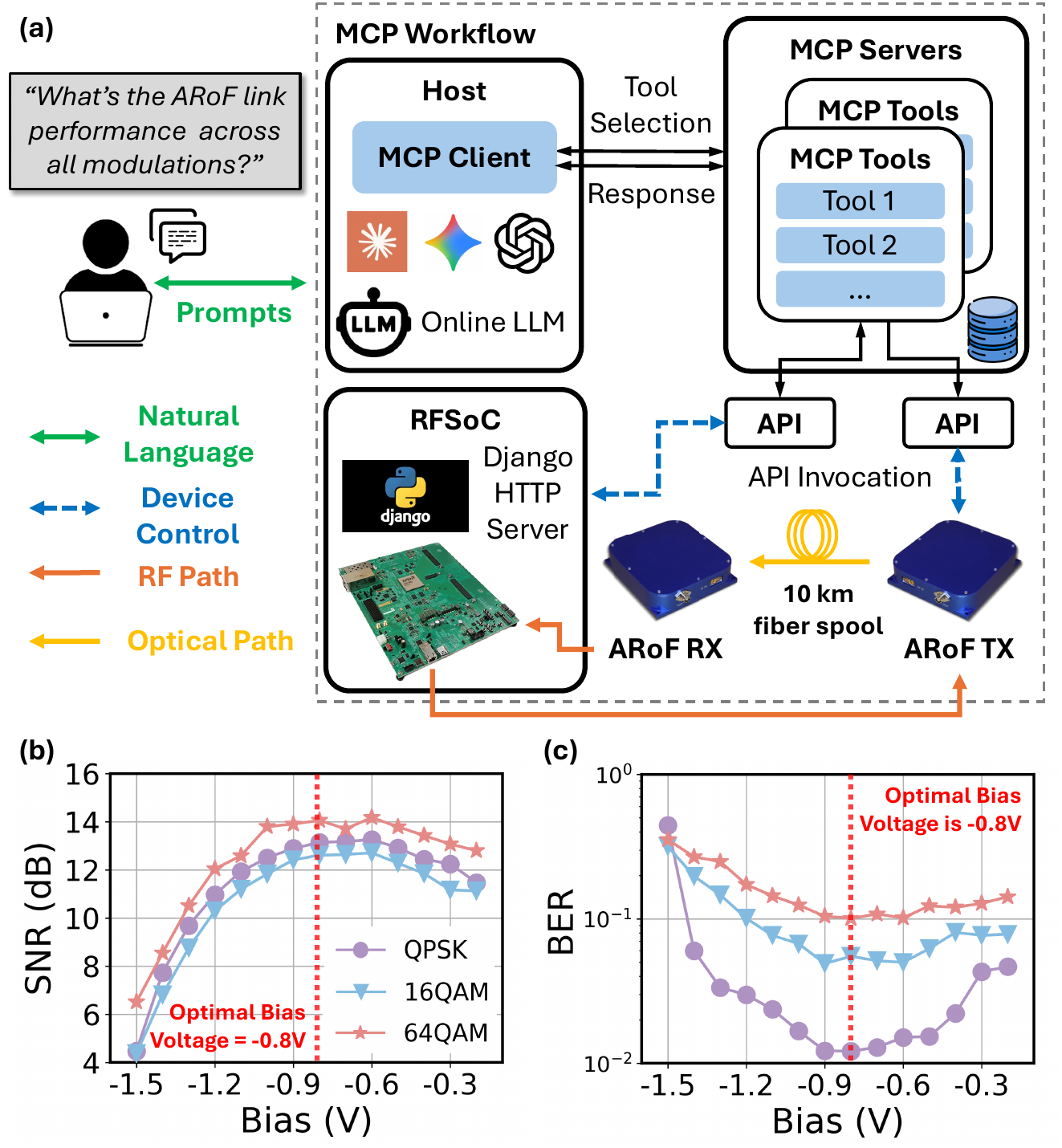}
    \caption{(a) {\name} workflow for LLM-assisted wide-bandwidth ARoF 5G new radio (NR) link with an RFSoC ZCU216 board and an ARoF transmitter-receiver pair.
    (b) and (c) Optimized ARoF transmitter bias voltage across link SNR and BER with different modulation orders, where the vertical dashed line indicates the optimized bias voltage selected by {\name}.}
    \label{fig:demo5}
    \vspace{-3.0mm}
\end{figure}

In the third case study, we demonstrate {\name} controlling a wireless-optical testbed to autonomously characterize an initially unknown transmitter configuration.
The experimental setup is shown in Fig.~\ref{fig:demo5}(a), where a radio frequency system-on-chip (RFSoC) ZCU216 board serves as the radio, generating and receiving a {400}\thinspace{MHz} orthogonal frequency-division multiplexing (OFDM) signal centered at {600}\thinspace{MHz} carrier frequency.
We adopt the RFSoC implementation from~\cite{wei2025spear} as the base hardware design, and the generated RF signal is subsequently modulated onto an optical carrier at {1552.44}\thinspace{nm} ({193.11}\thinspace{THz}) using an Optilab EAM.

Within this framework, {\name} acts as the orchestration layer, dynamically configuring and sweeping the ARoF transmitter bias voltage while continuously measuring the ARoF link SNR and BER.
This case study involves a measurement-driven workflow in which {\name} coordinates transmitter configuration, waveform testing, and link-quality measurement across the ARoF transmitter and RFSoC testbed. Based on the operator's characterization goal, {\name} invokes the corresponding MCP tools to sweep the specified bias-voltage range, test multiple modulation formats, collect SNR, EVM, and BER measurements, and identify the operating condition that provides improved wireless transmission performance.
Specifically, {\name} autonomously orchestrates the RFSoC ZCU216 board to transmit and receive 5G NR OFDM waveforms using quadrature phase-shift keying (QPSK) and 16- and 64-level quadrature amplitude modulation (16QAM and 64QAM), with onboard signal processing used to compute feedback metrics including SNR, EVM, and BER.
The modulated ARoF signal is propagated over a {10}\thinspace{km} fiber spool under varying ARoF transmitter bias voltages.
Figs.~\ref{fig:demo5}(b) and (c) show the measured SNR and BER of the received OFDM signal across the automated ARoF link configurations.
The results show that {\name} can characterize the ARoF link, identify, and apply the optimal ARoF transmitter bias voltage for improved wireless transmission performance.

\subsection{Adaptive Channel Configuration and generalized signal-to-noise ratio (GSNR) Optimization on Multi-Span Optical Links}

\begin{figure}[!t]
    \centering
    \includegraphics[width=0.98\columnwidth]{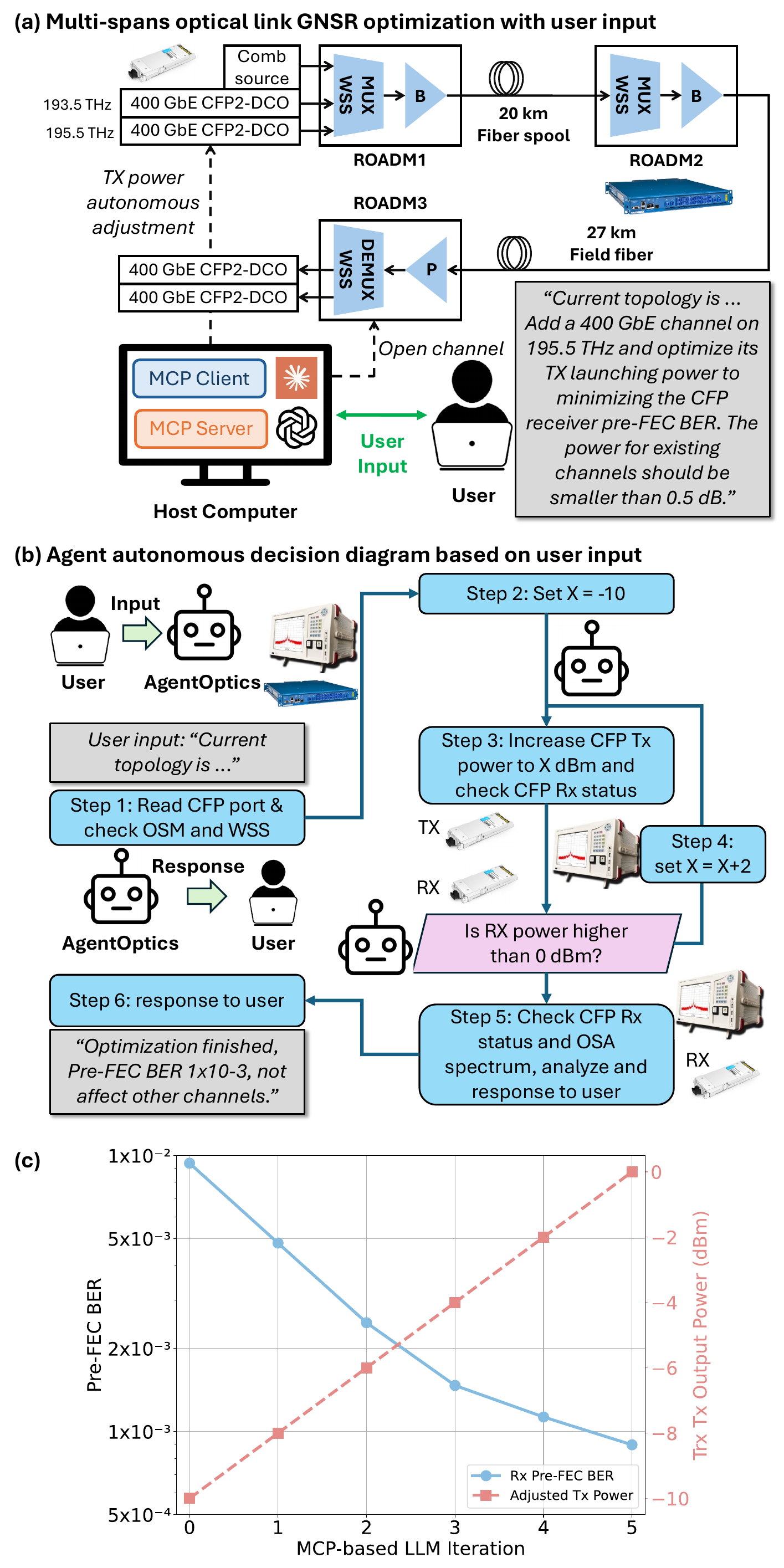}
    \caption{(a) {\name} provisions a 400\thinspace{GbE} channel in a two-span link and autonomously optimizes the channel GSNR based on a single-line human language instruction.
    (b) Autonomous launch power optimization of the CFP2-DCO 400\thinspace{GbE} transmitter (Tx) performed by {\name}.
    (c) Pre-FEC BER optimization of the 400\thinspace{GbE} signal by {\name} using an online LLM (Sonnet 4.5) without impacting existing background traffic.}
    \label{fig:diagram_demo2}
    \vspace{-3.0mm}
\end{figure}


For the fourth case study, we demonstrate that {\name} is capable not only of basic link establishment but also of reasoning-driven operations such as channel power optimization.
Rather than providing a sequence of device commands, the user specifies only the network topology and a high-level optimization objective. {\name} autonomously determines the required workflow by selecting the appropriate MCP tools, coordinating multiple optical devices, and iteratively optimizing the transmitter launch power based on measurement feedback until the optimization objective and operational constraints are satisfied.
To illustrate this capability, we establish a two-span optical link consisting of two 400\thinspace{GbE} CFP2-DCO transceivers and an amplified spontaneous emission (ASE) comb source, as shown in Fig.~\ref{fig:diagram_demo2}.
Specifically, ten ASE channels with {50}\thinspace{GHz} channel spacing from {193.85 to 194.85}\thinspace{THz} are injected at the first ROADM to emulate background traffic.
At ROADM1, the WSS multiplexes the ASE channels and two 400\thinspace{GbE} signals from CFP2-DCOs, which are subsequently amplified by a booster EDFA and transmitted over a {20}\thinspace{km} fiber spool.
The signals then pass through ROADM2 and subsequently propagate over a {27}\thinspace{km} field fiber.
Finally, the composite signal is amplified by a pre-amplifier and demultiplexed by the WSS in ROADM3, where the background ASE channels are dropped, and the two 400\thinspace{GbE} channels are directed to the receiver CFP2-DCO.

In this case study, we demonstrate {\name}'s capability for real-time dynamic link reconfiguration.
Specifically, it adds a new 400\thinspace{GbE} channel centered at {195.5}\thinspace{THz} to the existing two-span link using a CFP2-DCO, and optimizes the transmitter launch power in order to minimize the pre-FEC BER at the receiver to be below a given threshold.
The user inputs link topology and reconfiguration requirements, including channel adjustment constraints, minimal impact on existing channels, and 400\thinspace{GbE} channel optimization objectives, into {\name}, as shown in Fig.~\ref{fig:diagram_demo2}(a).
{\name}'s behavior during the channel power optimization is shown in Fig.~\ref{fig:diagram_demo2}(b), which is determined by {\name} itself. Note that we do not provide a script for {\name} for optimization, just providing basic device API-level MCP tools. The user specifies the topology and optimization objective. In Step~1, {\name} reads CFP port parameters and checks OCM and WSS status. In Step~2, the CFP Tx power is initialized to $X=-10$\thinspace{dBm}. In Step~3, {\name} increases CFP Tx power to $X$\thinspace{dBm} and queries CFP transmitter (Tx)/ receiver (Rx) status. A decision block evaluates whether CFP Rx power exceeds 0\thinspace{dBm}; if so, Step~4 updates the Tx power to $X+2$\thinspace{dB} and repeats the loop. In Step~5, {\name} analyzes CFP Rx status and OSA spectrum to verify pre-FEC BER and inter-channel impact. Finally, in Step~6, the agent reports the optimization results to the user, confirming minimized pre-FEC BER and negligible impact on existing channels.

The corresponding optimization results are shown in Fig.~\ref{fig:diagram_demo2}(c), where the $x$-axis denotes the iteration index as {\name} sequentially invokes different MCP tools to fulfill the user request for optimizing the BER for the newly added 400\thinspace{GbE} channel.
A series of iterative steps are performed with a 2\thinspace{dB} step size to optimize the launch power of the CFP2-DCO, which is autonomously determined by the LLM (see Fig.~\ref{fig:diagram_demo2}(b), right $y$-axis).
The left $y$-axis depicts the evolution of the pre-FEC BER of the newly added channel, demonstrating convergence to a minimized BER upon completion of the LLM-based optimization process.
The optimization process automatically terminates once the CFP2-DCO receiver power reaches its predefined threshold.

\subsection{Polarization Monitoring and Stabilization}

The fifth case study demonstrates agentic orchestration for automated polarization stabilization, which is crucial for a wide range of applications, including coherent optical communication~\cite{ji2023selfcoherentreceiver}, distributed interferometric fiber sensing~\cite{fu2022polfadingdas}, and polarization-sensitive quantum links such as entanglement-based quantum key distribution (QKD)~\cite{yin2025polcomp}.
Maintaining polarization stability is essential for fiber-optic links because environmental perturbation, such as temperature variation and vibration, can induce polarization drift over time and cause performance degradation.
Traditionally, mitigating this drift requires researchers to manually activate polarization stabilization procedures upon observing degradation.
In contrast, {\name} enables polarization correction through a single natural-language command, greatly simplifying operation and reducing response time.

\begin{figure}[!t]
    \centering
    \includegraphics[width=0.98\columnwidth]{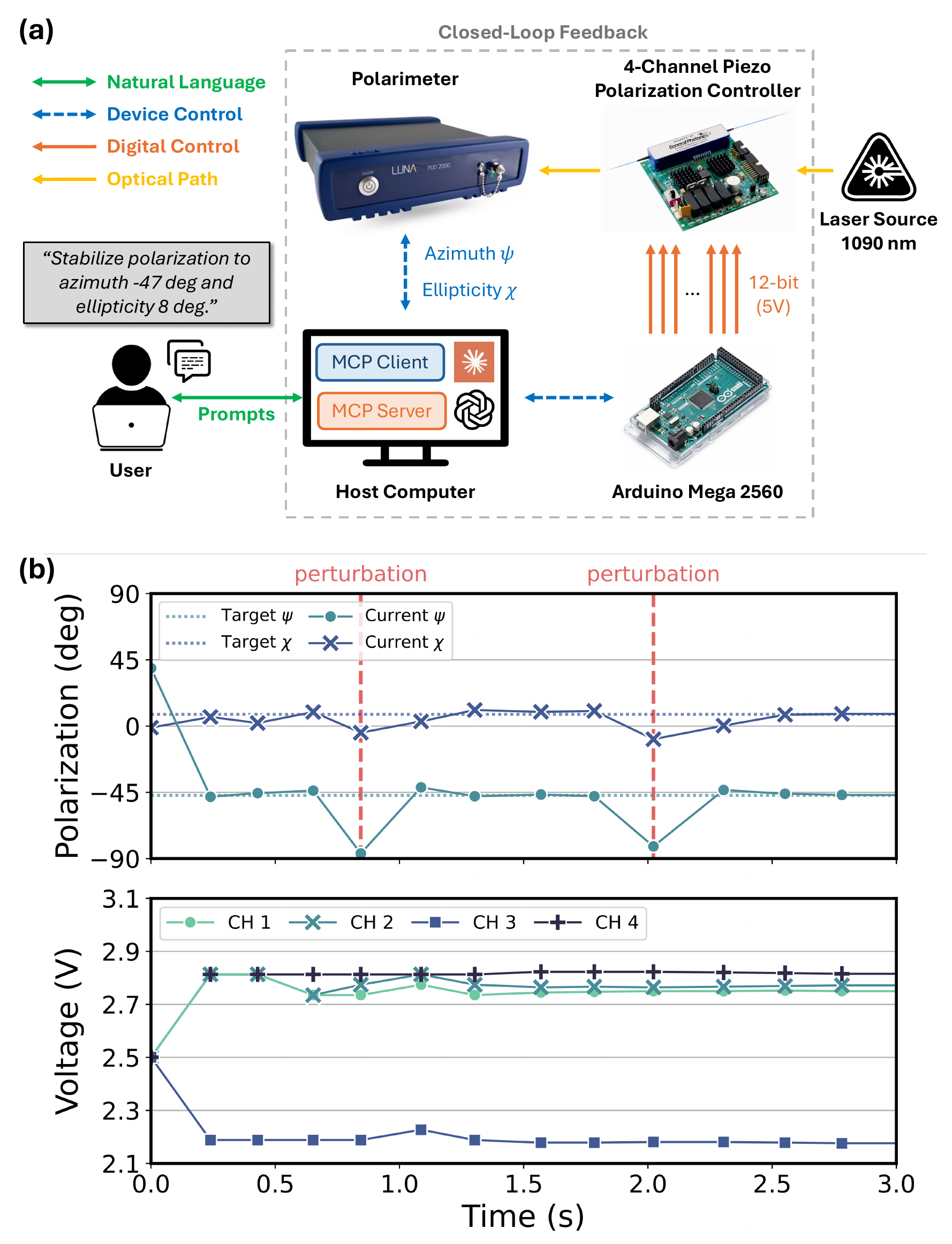}
    \vspace{-2mm}
    \caption{(a) Experimental setup and control architecture for fiber link polarization stabilization using {\name}.
    (b) Closed-loop polarization stabilization results with deliberate fiber perturbations, showing polarization state and piezo controller actuation over time. Red dashed vertical lines indicate the perturbation times.}
    \label{fig:diagram_demo3}
    \vspace{-3.0mm}
\end{figure}

The experimental setup, as shown in Fig.~\ref{fig:diagram_demo3}(a), consists of a 1090\thinspace{nm} laser source, a Luna PCD-M02 4-channel piezo polarization controller, and a Luna POD2000 polarimeter.
A host computer running an MCP client interfaces with an MCP server, which communicates with an Arduino Mega 2560 via USB.
The Arduino device drives the PCD-M02 controller using 12-bit digital control codes from 0 to 4095 via a digital-to-analog converter (DAC) interface; these codes map to an output voltage range from {0 to 5}\thinspace{V} with a step size of {1.22}\thinspace{mV}.
In this case study, {\name} orchestrates two MCP-controlled devices, the POD2000 for polarization state measurement and the PCD-M02 via Arduino for actuation, and completes the workflow by invoking four tools: configure the POD2000 wavelength (1090\thinspace{nm}), read the current polarization state, stabilize to azimuth $-${47}\thinspace{deg} and ellipticity 8\thinspace{deg} with stopping threshold of {0.5}\thinspace{deg}, and re-read the polarization state for verification.
The LLM parses the target azimuth, ellipticity, wavelength, and stopping threshold from natural language and maps them onto the corresponding MCP tool arguments. It also reasons over the device coupling, e.g., issuing the wavelength configuration on the polarimeter before readout and scheduling a post-stabilization readout for verification, so that different operator inputs yield correspondingly adapted tool invocations rather than a fixed script.
The polarization stabilization is implemented using a multi-stage gradient descent procedure that iteratively adjusts the four piezo control codes while monitoring the polarization state until convergence criteria are met. 

Fig.~\ref{fig:diagram_demo3}(b) shows the results given the input prompt, including a series of deliberate fiber perturbations to demonstrate system robustness.
Starting from a random initial state far from the target ($\psi=39.33^{\circ}$, $\chi=-1.05^{\circ}$), the system converges to target values ($\psi=-47^{\circ}$, $\chi=8^{\circ}$), achieving the stopping criterion (angular error $<0.5^{\circ}$) after 12 iterations, with an average iteration time of {0.23}\thinspace{sec}.
For each optimization iteration, the controller performs multiple polarimeter readings to decide the actuation direction: for each piezo channel, it applies one small step forward and one small step backward, measures the resulting polarization states, and executes an actuation step in the direction that most significantly reduces the angular error relative to the target.
During convergence, the system successfully recovers from multiple perturbations, including manual fiber disturbances that temporarily shift the polarization by over 40$^{\circ}$ (visible at iterations 4 and 9).
The piezo control voltages display smooth optimization trajectories with significant adjustments across all channels: starting from the initialization at 2.5\thinspace{V}, channels 1, 2, and 4 increase to {2.75}\thinspace{V}, {2.77}\thinspace{V}, and {2.81}\thinspace{V} respectively, while channel 3 decreases slightly to {2.18}\thinspace{V} during the optimization process. 

\subsection{DAS-enabled Fiber Sensing and Event Detection}

\begin{figure}[!t]
    \centering
    \includegraphics[width=0.98\columnwidth]{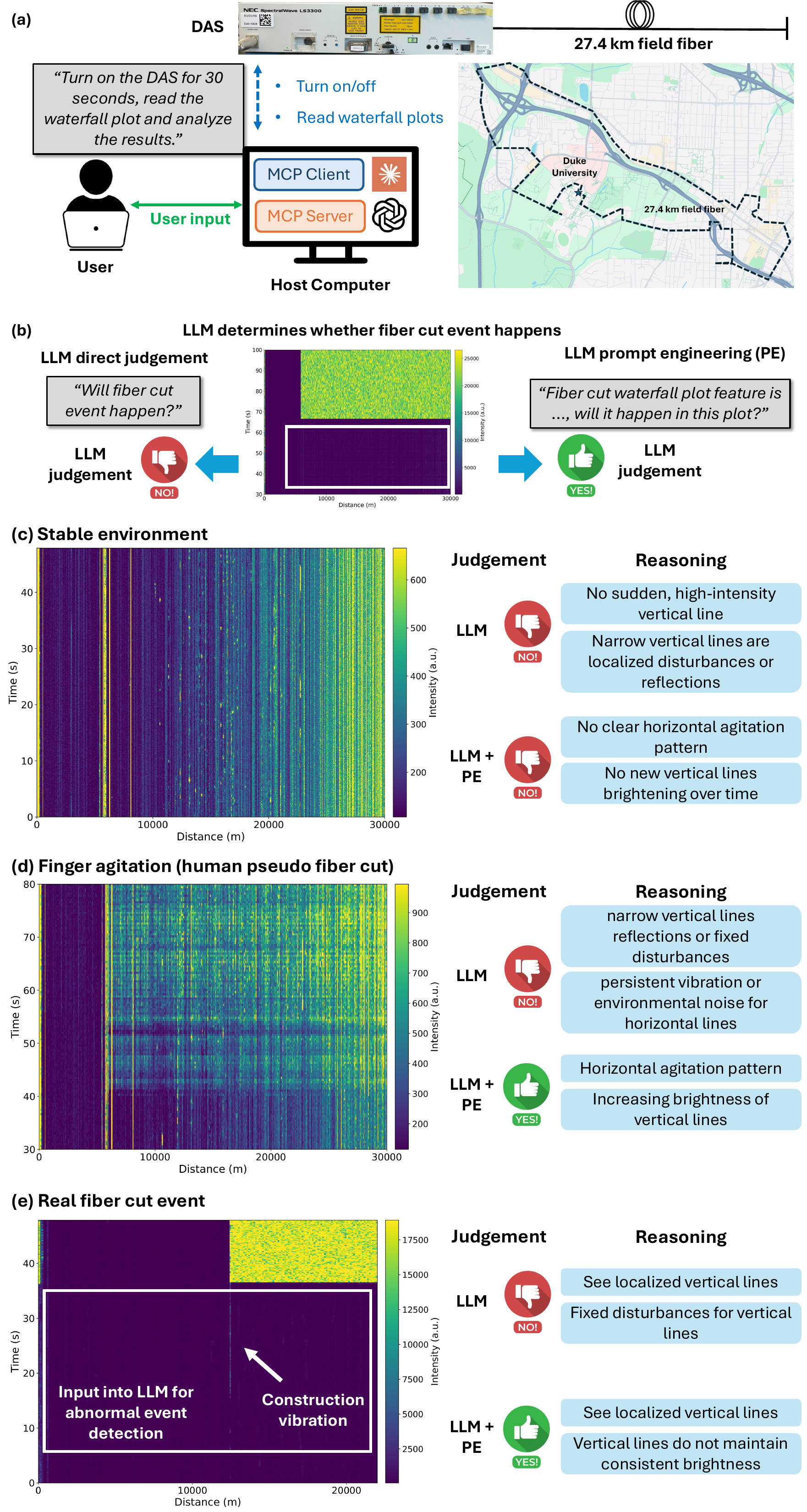}
    \caption{(a) Experimental setup and workflow for {\name}-enabled fiber monitoring using DAS.
    (b) LLM-based reasoning and prompt engineering (PE) for automated event interpretation on the DAS waterfall plot analysis for (c) a stable environment, (d) human-induced pseudo fiber agitation, and (e) a real fiber cut event.}
    \label{fig:diagram_demo4}
    \vspace{-3.0mm}
\end{figure}

In the sixth case study, we show that {\name} can also assist in monitoring fiber conditions using DAS and in determining whether a potential fiber cut event may occur that could affect optical link transmission performance. Current DAS image analysis typically requires scenario-specific machine learning methods, leading to the need for specialized and often complex algorithm design for each case. We demonstrate that agentic-based DAS control and image analysis can be achieved using only a few natural language instructions, even for complex fiber cut event detection.
To support an agentic workflow for DAS operation, we develop MCP tools for an NEC Spectral LS3300 DAS interrogator, which is set up to monitor a 27.4\thinspace{km} field fiber, as shown in Fig.~\ref{fig:diagram_demo4}(a).
The MCP tools support DAS vibration monitoring over a specified time window and retrieval of the corresponding waterfall plot images.
A waterfall plot shows vibration intensity along the sensing fiber over time, with distance along the fiber on the horizontal axis, time on the vertical axis, and color indicating vibration amplitude or strain rate.
This representation enables rapid identification of the location, timing, and strength of vibration events.
The returned images are autonomously analyzed by an LLM to determine potential fiber cut events based on abnormal vibration patterns. The LLM is responsible for interpreting the waterfall plot and deciding subsequent workflow actions, whereas DAS data acquisition and signal generation are performed by dedicated MCP tools.

Without domain-specific guidance, {\name} often fails to accurately identify fiber cut events due to limited knowledge of fiber sensing characteristics.
To address this challenge, we apply prompt engineering~\cite{shin2020autoprompt} to supply additional domain knowledge related to fiber-sensing waterfall plots, thereby enhancing the reasoning capability of {\name} for DAS analysis.
In addition to requesting a determination of whether a given waterfall plot indicates an impending fiber cut, we provide background information in the prompt describing characteristic pre-cut waterfall plot patterns. 
This information is combined with the fiber cut determination request and the corresponding waterfall plot and fed into the LLM.
We further validate this prompt engineering approach for fiber cut prediction using a previously recorded real-world fiber cut event that occurred on a {53}\thinspace{km} field-deployed fiber.
Figs.~\ref{fig:diagram_demo4}(c) through (e) compare the LLM-based waterfall analysis with and without prompt engineering under three scenarios:
a stable environment,
a slight fiber perturbation event induced by manually agitating the fiber to generate horizontal bright lines in the waterfall plot, and
a real fiber cut event recorded on a {53}\thinspace{km} field fiber loop.
The prompt provides explicit descriptions of fiber cut signatures in waterfall plots, such as: ``\textit{A fiber cut event is detected when fiber agitation produces multiple vertical streaks, or when unequal brightness between the top and bottom vertical lines indicates a power discontinuity}.''
As shown in Figs.~\ref{fig:diagram_demo4}(c) through (e), incorporating prompt engineering enables the LLM to reliably detect fiber cut events, thereby demonstrating that {\name} can support an MCP-driven fiber sensing and monitoring workflow with LLM-assisted event interpretation to help identify potential fiber cut events.



\section{Conclusions}
\label{sec:conclusions}

In this paper, we presented {\name}, a language-driven orchestration layer built on MCP that provides a standardized interface between LLM reasoning and heterogeneous optical device APIs for autonomous and scalable optical system operation, and systematically evaluated its performance through a comprehensive benchmark on real optical hardware and six representative case studies.
{\name} implements 64 standardized tools across eight physical devices and is systematically evaluated using a 410-task benchmark conducted on real hardware.
We assess performance using four commercial online LLMs and five locally hosted Gemma and Llama models, and compare against the direct LLM-based code generation baselines.
Overall, {\name} achieves average task success rates of 98.8\% with online models and 65.4\% across the local Gemma and Llama models. The three Gemma models remain robust, with success rates from 79.8\% to 84.4\%; Llama-3.1-8B reaches 71.5\%, and Llama-3.2-3B reaches 7.1\%. All local results use unquantized checkpoints.
We further validate {\name} through five representative runtime case studies, ranging from DWDM signal provisioning and monitoring to DAS-based fiber sensing with LLM-assisted event interpretation. We also demonstrate an LLM-assisted workflow for generating MCP-compatible tool interfaces from legacy device-control code, supporting scalable onboarding of additional optical device capabilities.
These case studies collectively demonstrate {\name}'s ability to coordinate heterogeneous optical devices, execute closed-loop optimization, and enable intelligent orchestration in real-world testbeds.
For practical deployment, LLM-based automation must still operate within the engineering and operational constraints of physical optical systems. Each new device may require vendor-specific backend integration, some configuration actions require topology-aware recovery rather than simple rollback, and operational networks usually require access-control and audit policies before autonomous write operations are allowed. Therefore, {\name} should be viewed as a structured orchestration layer built on validated device tools, rather than a complete replacement for site-specific engineering and network-operation policies.
Similarly, {\name} is not intended to replace domain-specific signal-processing, QoT prediction, or physical-layer models; instead, such measurement, analysis, and prediction modules can be integrated as MCP tools, with {\name} coordinating their use for monitoring, optimization, and closed-loop control.
Future work will expand the {\name} MCP toolset to incorporate broader classes of optical and hybrid wireless-optical systems, improve robustness for large-scale and long-horizon orchestration tasks, and incorporate enhanced safety mechanisms and cost-efficient deployments.
We also plan to enhance {\name} in more complex operational environments to further advance autonomous optical network control and management.

\section*{Appendix}

\subsection{DWDM Link Setup and Performance Monitoring}

\noindent\textbf{Prompt}: ``\textit{Add a WSS connection named `ARoF' with frequency from 193900\thinspace{GHz} to 194450\thinspace{GHz} with attenuation of 19.5\thinspace{dB} on the MUX side, using input port 1 and connection ID 1.
Add a WSS connection named `CFP2' with frequency from 193400\thinspace{GHz} to 193700\thinspace{GHz} with attenuation of 5\thinspace{dB} on the MUX side, using input port 20 and connection ID 2. 
Set the OSA start wavelength to 1540\thinspace{nm} and stop wavelength to 1550\thinspace{nm}.
Add a WSS connection named `ARoF' from 193900\thinspace{GHz} to 194450\thinspace{GHz} with attenuation of 5\thinspace{dB} on the DEMUX side, using output port 1 and connection ID 1. 
Add a WSS connection named `CFP2' from 193400\thinspace{GHz} to 193700\thinspace{GHz} with attenuation of 5\thinspace{dB} on the DEMUX side, using output port 20 and connection ID 2. 
Set the CFP2 frequency to 193500000\thinspace{MHz}. 
Set the ARoF Tx current to 99\thinspace{mA} and bias voltage to $-$0.9\thinspace{V}.}''

\subsection{Wideband 5G ARoF Link Characterization}

\noindent\textbf{Prompt}: ``\textit{You are an automated experiment controller tasked with finding the optimal bias voltage for an analog radio-over-fiber (ARoF) link.
The testbed consists of an RFSoC that generates and receives RF signals routed through optical transceivers, with system parameters fixed as follows: RF bandwidth at {400}\thinspace{MHz}, DAC NCO at {600}\thinspace{MHz}, ADC NCO at $-${600}\thinspace{MHz}, RF attenuation at {0}\thinspace{dB}, and bias current at {99}\thinspace{mA}.
Using the {\name} MCP server, sweep the bias voltage from $-${1.5}\thinspace{V} to {0}\thinspace{V} in steps of {0.1}\thinspace{V}, and at each bias point use the `rfsoc link tester' MCP server to measure EVM, SNR, and BER for three modulation schemes: QPSK, 16QAM, and 64QAM.
Once all measurements are collected, generate SNR-vs-bias and BER-vs-bias plots for all modulation schemes, and write the results into three markdown files, each organized as a table with bias voltage and per-modulation metric values clearly presented.}''

\subsection{Adaptive Channel Configuration and GSNR Optimization on Multi-Span Optical Links}

\noindent\textbf{Prompt}: ``\textit{Set a new CFP channel output power to -10\thinspace{dBm}, add an ROADM 2 WSS single connection for a 400\thinspace{GbE} channel at 195.5\thinspace{THz}, and record the CFP input power and pre-FEC BER. The current link is CFP transmitter $\rightarrow$ ROADM 1 booster $\rightarrow$ 20\thinspace{km} fiber $\rightarrow$ ROADM 2 booster $\rightarrow$ 27\thinspace{km} field fiber $\rightarrow$ ROADM 3 preamp $\rightarrow$ CFP receiver. Optimize the CFP launching power from $-$15 to 0\thinspace{dBm} to minimize the pre-FEC BER of the added CFP channel without adjusting any ROADM gains, ensuring the received power of the new CFP channel remains below 0\thinspace{dBm}. An existing CFP channel is already stable; its pre-FEC BER must not change significantly during optimization, and the ROADM 2 demux input power for channels 1 to 20 must not vary by more than $\pm$0.5\thinspace{dB} before and after optimization. Record each step, including pre-FEC BER for both the existing and newly added CFP channels, to determine the optimal launching power.}''

\subsection{Polarization Monitoring and Stabilization}

\noindent\textbf{Prompt}: ``\textit{Configure POD2000 at {1090}\thinspace{nm} wavelength, get the current polarization state, then stabilize the polarization to azimuth $-$47\thinspace{deg} and ellipticity 8\thinspace{deg} with a stop threshold of 0.5\thinspace{deg}, and verify the result.}''

\subsection{DAS-Enabled Fiber Sensing and Event Detection}

\noindent\textbf{Prompt}: ``\textit{A fiber cut event can occur under two conditions: (1) fiber agitation, which appears as horizontal lines in the waterfall plot even when they are not dominant; this agitation is often accompanied by multiple vertical lines, indicating a potential fiber cut event; and (2) a mismatch in brightness between the vertical lines at the top and bottom of the plot, which may also suggest a fiber cut. Based on these criteria, does this waterfall plot indicate a fiber cut event?}''





\bibliographystyle{IEEEtran}
\bibliography{reference}


\begin{IEEEbiography}[{\includegraphics[width=1in,height=1.25in,clip,keepaspectratio]{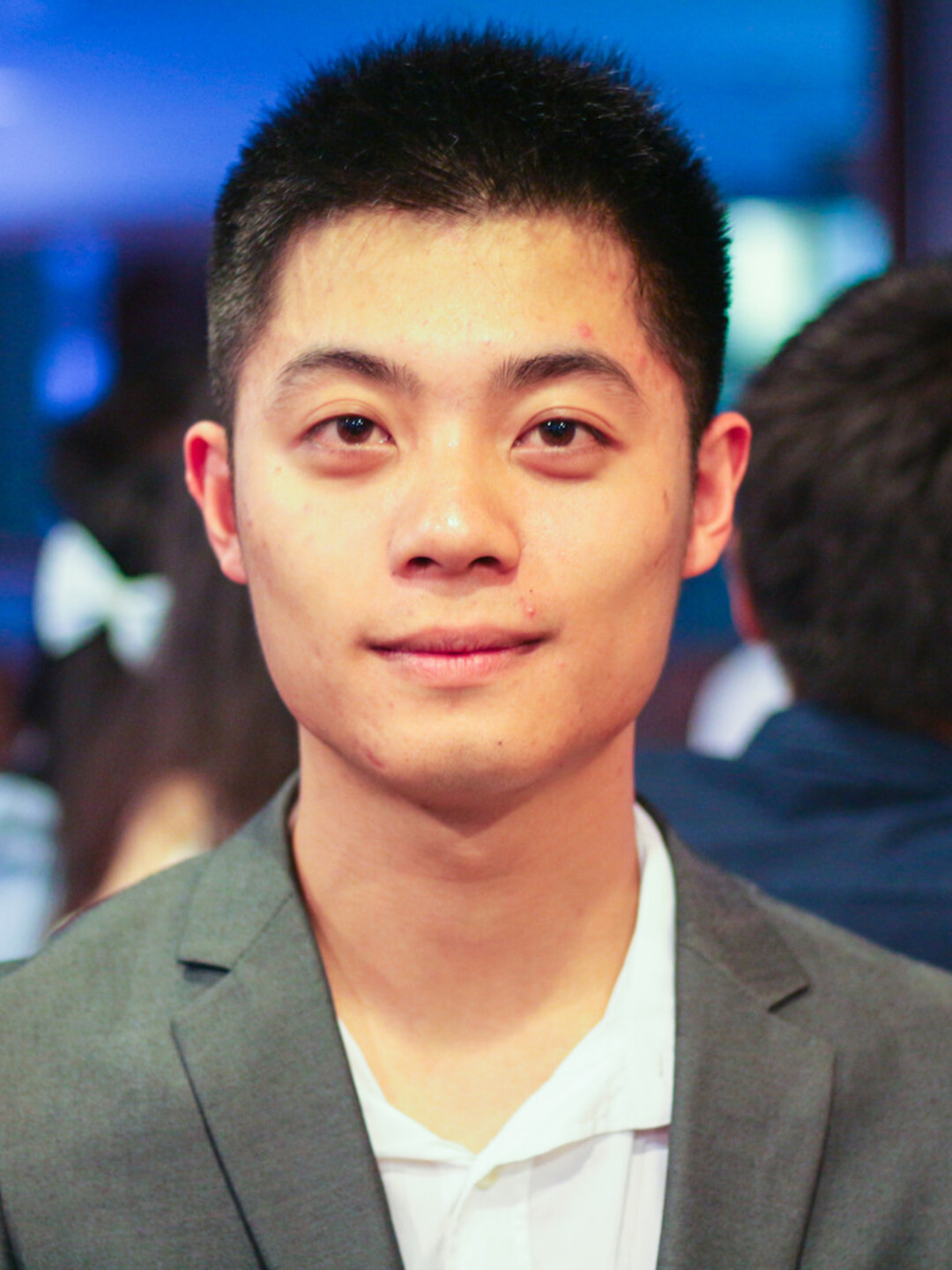}}]{Zehao Wang}~is currently pursuing his Ph.D. degree in the Department of Electrical and Computer Engineering at Duke University. He received his B.Eng. degree in information science from Zhejiang University, China, in 2019, and an M.S. degree in Electrical and Computer Engineering from Carnegie Mellon University in 2020. His research interests are in the areas of next-generation optical-wireless-quantum networks with a focus on integrated sensing and communication systems and radio-over-fiber technologies. He was selected as a Corning Outstanding Student Paper Competition finalist at OFC’24, and received several top-scored papers in OFC.
\end{IEEEbiography}

\begin{IEEEbiography}[{\includegraphics[width=1in,height=1.25in,clip,keepaspectratio]{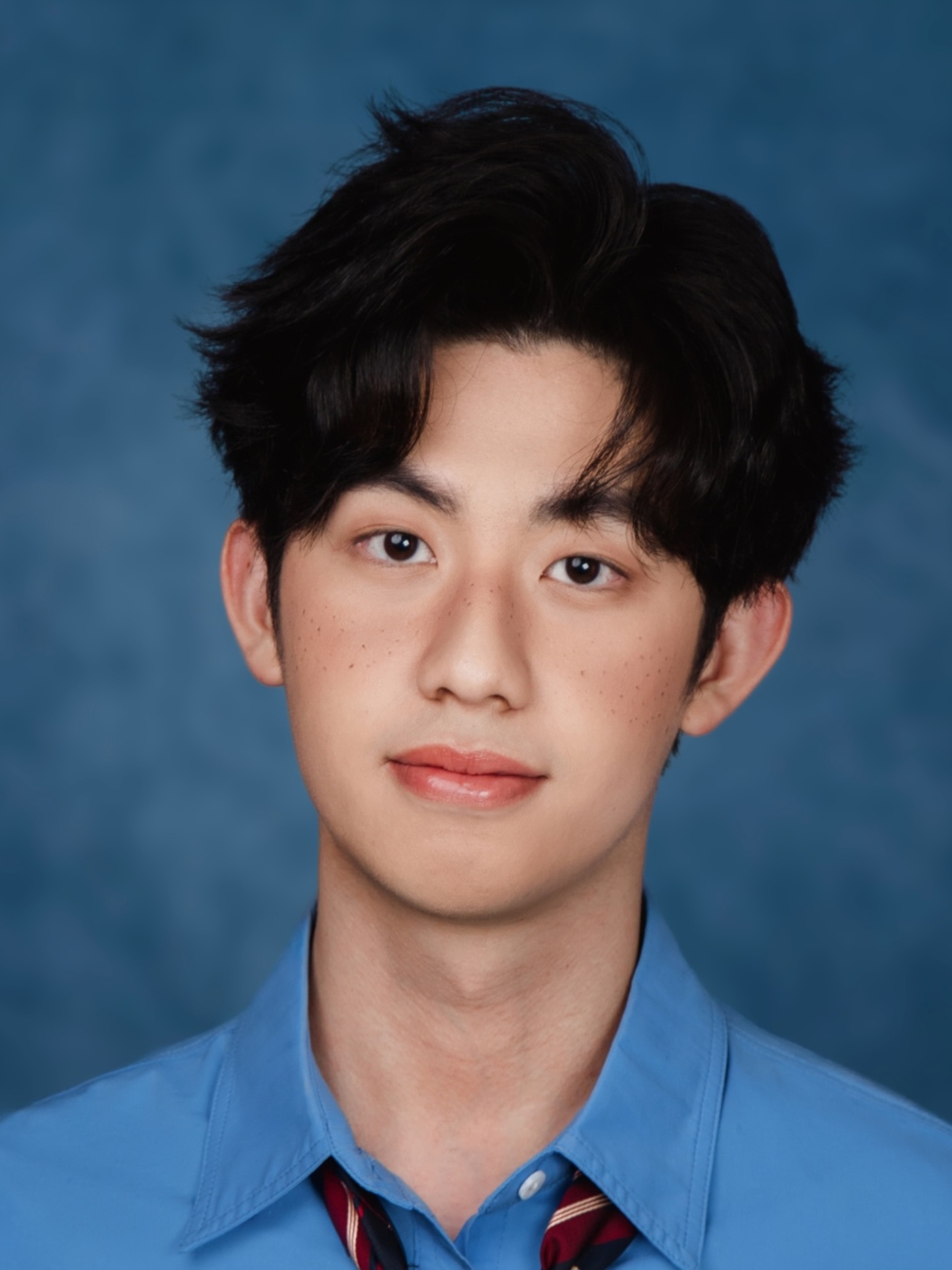}}]{Mingzhe Han}~is currently pursuing his Ph.D. degree in the Department of Electrical and Computer Engineering at Duke University. He received his B.S. degree in Optical Engineering and a B.A. degree in Computer Science from the University of Rochester, and completed an M.S. program in Electrical Engineering at California Institute of Technology (Caltech). His research interests are in the areas of optics and networking, and he also works on machine learning and agentic systems.
\end{IEEEbiography}

\begin{IEEEbiography}[{\includegraphics[width=1in,height=1.25in,clip,keepaspectratio]{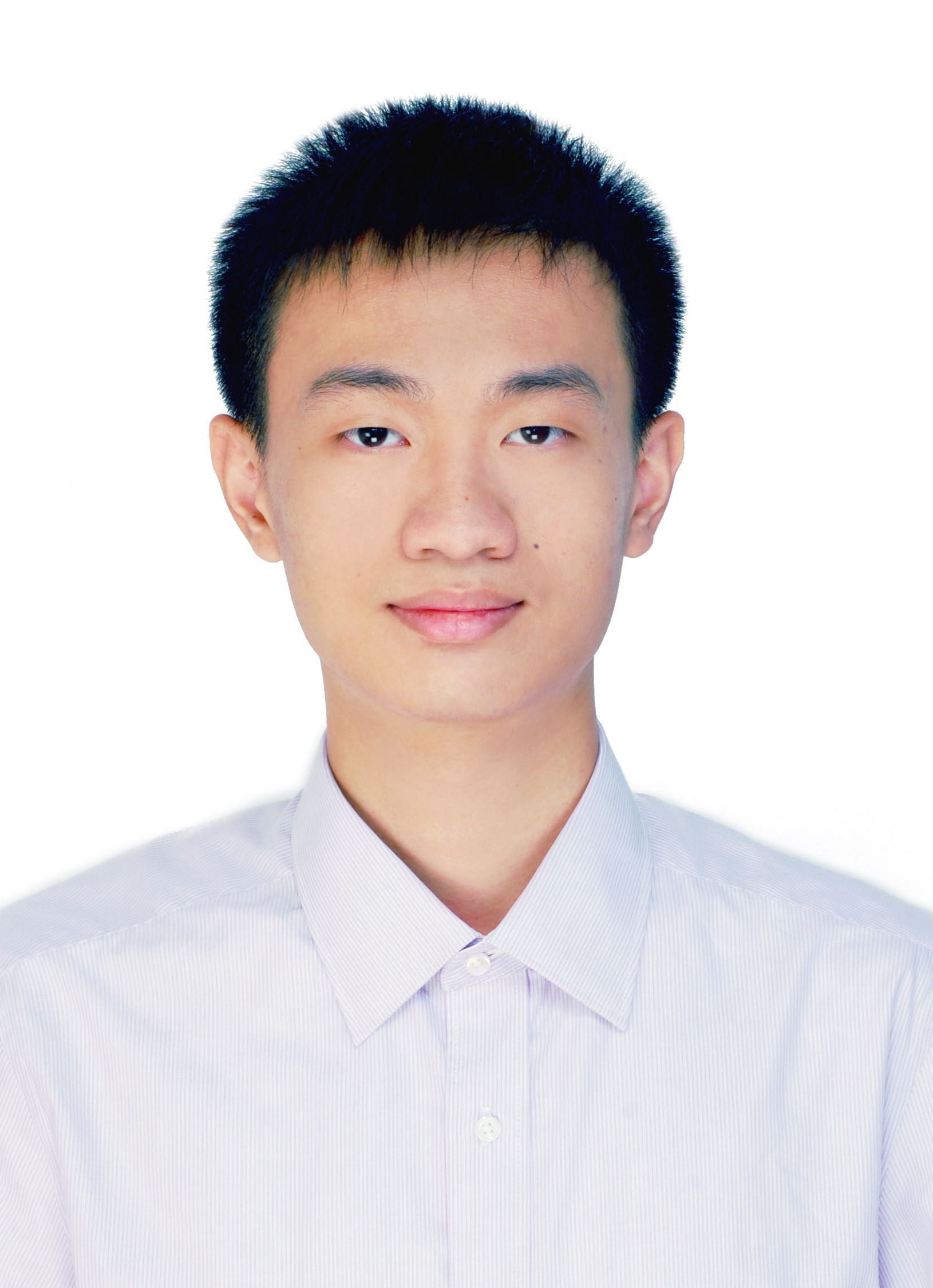}}]{Wei Cheng}~earned his B.E. in Computer Engineering from the University of Hong Kong in 2018 and an M.S. in Computer Science and Information Engineering from National Cheng Kung University in 2022. He is currently a Ph.D. student in Electrical and Computer Engineering at Duke University, where his research focuses on AI accelerators, computer architecture, communication, and quantum control systems.
\end{IEEEbiography}

\begin{IEEEbiography}[{\includegraphics[width=1in,height=1.25in,clip,keepaspectratio]{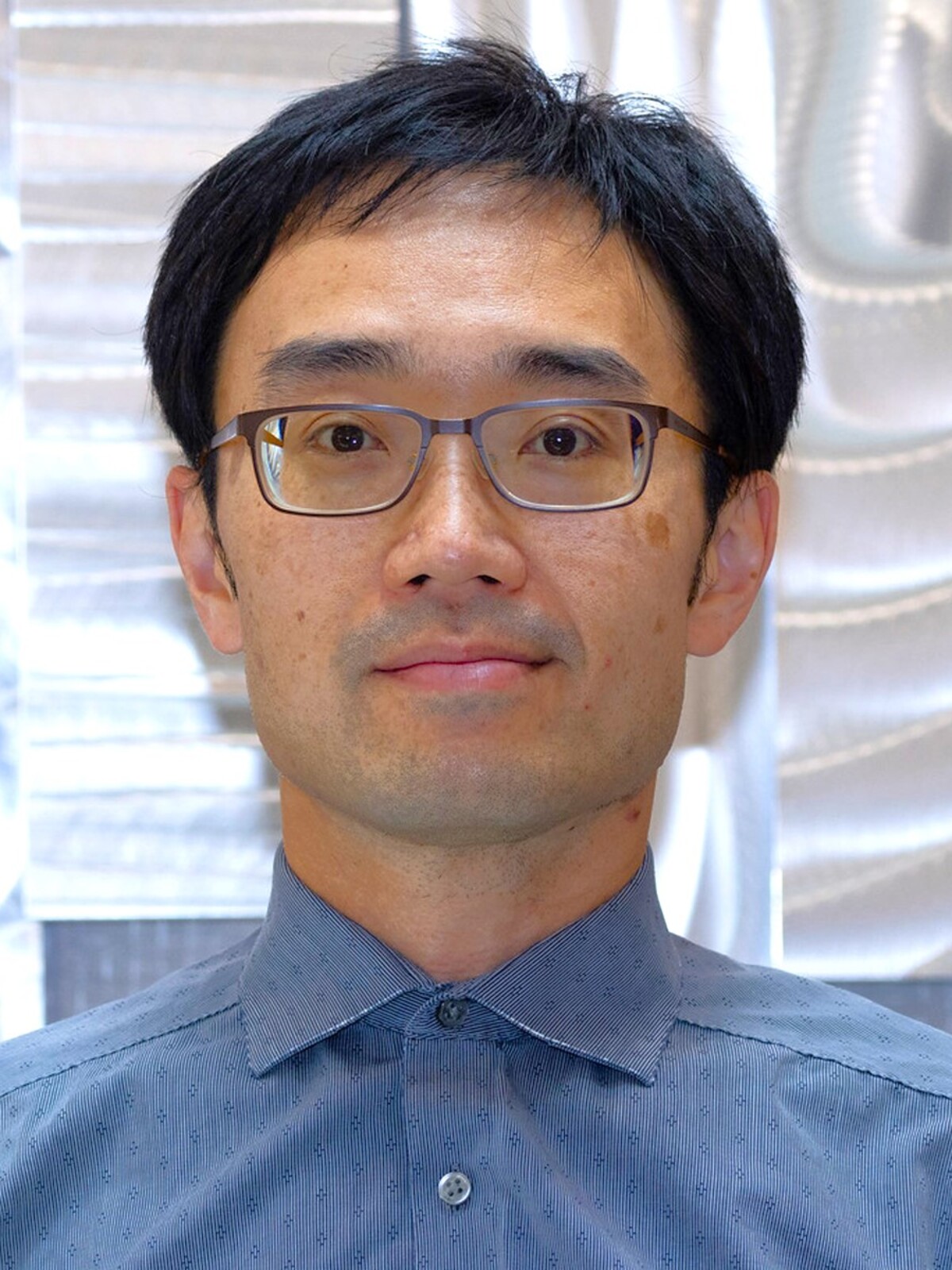}}]{Yue-Kai Huang}~is a Senior Researcher at NEC Laboratories America in Princeton, NJ. He received his PhD in Electrical and Electronics Engineering from Princeton University. At NEC, Dr. Huang’s work advances the field of optical networking and fiber-based sensing systems. His research includes long-distance fiber transmission, system design for distributed fiber sensing, and optical computation techniques using highspeed photonics. His contributions has resulted in many of NEC’s products in coherent WDM systems and DAS sensing solutions. He has published more than 100 journal and conference papers, and has been awarded with more than 40 US patents.
\end{IEEEbiography}

\begin{IEEEbiography}[{\includegraphics[width=1in,height=1.25in,clip,keepaspectratio]{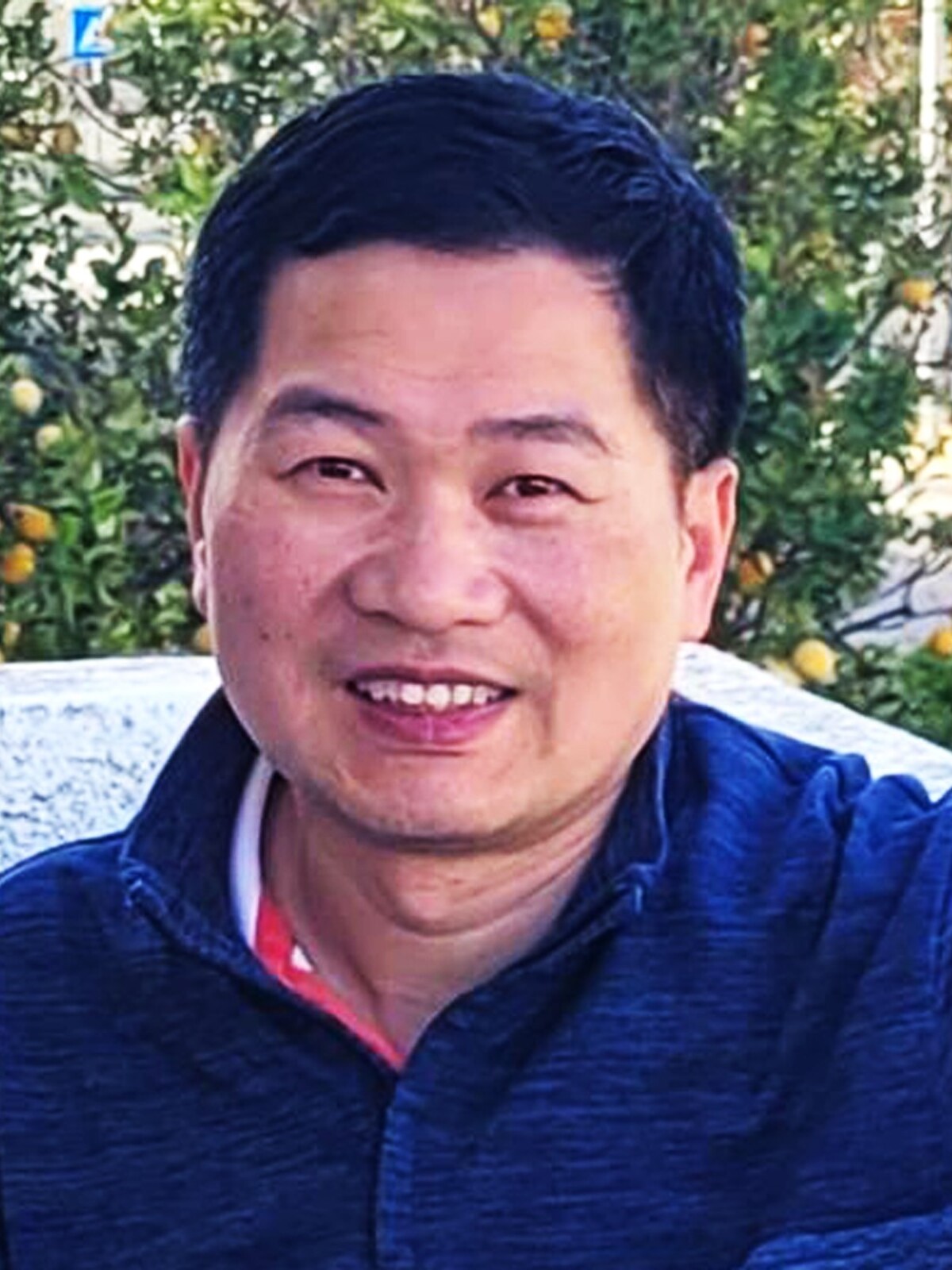}}]{Philip N. Ji}~is a Senior Researcher in the Optical Networking and Sensing Department at NEC Laboratories America, where he has been researching optical switching, networking, and sensing technologies since 2001. He received his B.E., M.E., and Ph.D. in Electrical Engineering \& Telecommunications from the University of New South Wales (UNSW), where he focused on special polymer optical fiber and fiber optic device innovation.
\end{IEEEbiography}

\begin{IEEEbiography}[{\includegraphics[width=1in,height=1.25in,clip,keepaspectratio]{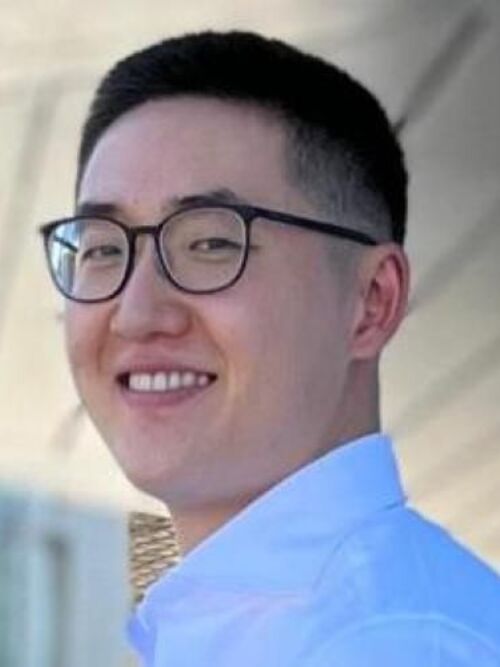}}]{Denton Wu}~completed a B.A. in physics at the University of Colorado Boulder. His undergraduate thesis focused on improving Pound-Drever-Hall laser locking robustness using microcontroller-automated monitoring and feedback. He then worked in the biomedical devices industry engineering field-deployable total internal reflection fluorescence microscopes. Denton began his PhD in physics at the University of Maryland in the group of Professor Norbert Linke and moved to Duke University with his advisor to finish his degree. His research focuses on building an ion trap quantum network node for testing novel entanglement generation schemes using the Strontium-88 ion. The research seeks to integrate ion-photon entanglement generation with existing telecom infrastructure, and investigates infidelity mechanisms affecting single photons sent through deployed fibers.
\end{IEEEbiography}

\begin{IEEEbiography}[{\includegraphics[width=1in,height=1.25in,clip,keepaspectratio]{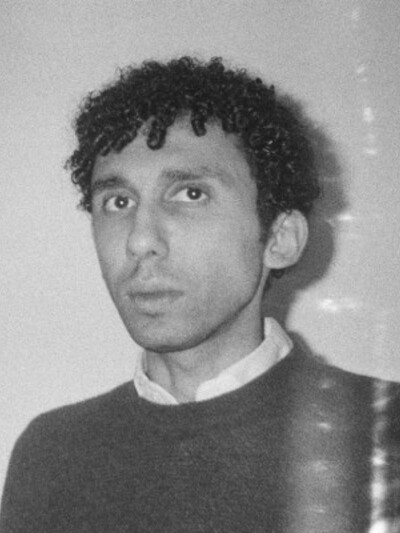}}]{Mahdi Safari}~is a computational design engineer at Axiomatic\_AI.  He received his M.S. in electrical and computer engineering at the University of Toronto. He has contributed to the development of advanced metasurfaces and metamaterials that exhibit both optical and radio-frequency transparency, with applications in smart communication systems and energy-efficient technologies. He specializes in the design and fabrication of novel nano/micro materials and their applications in radio frequencies and photonics. 
\end{IEEEbiography}

\begin{IEEEbiography}[{\includegraphics[width=1in,height=1.25in,clip,keepaspectratio]{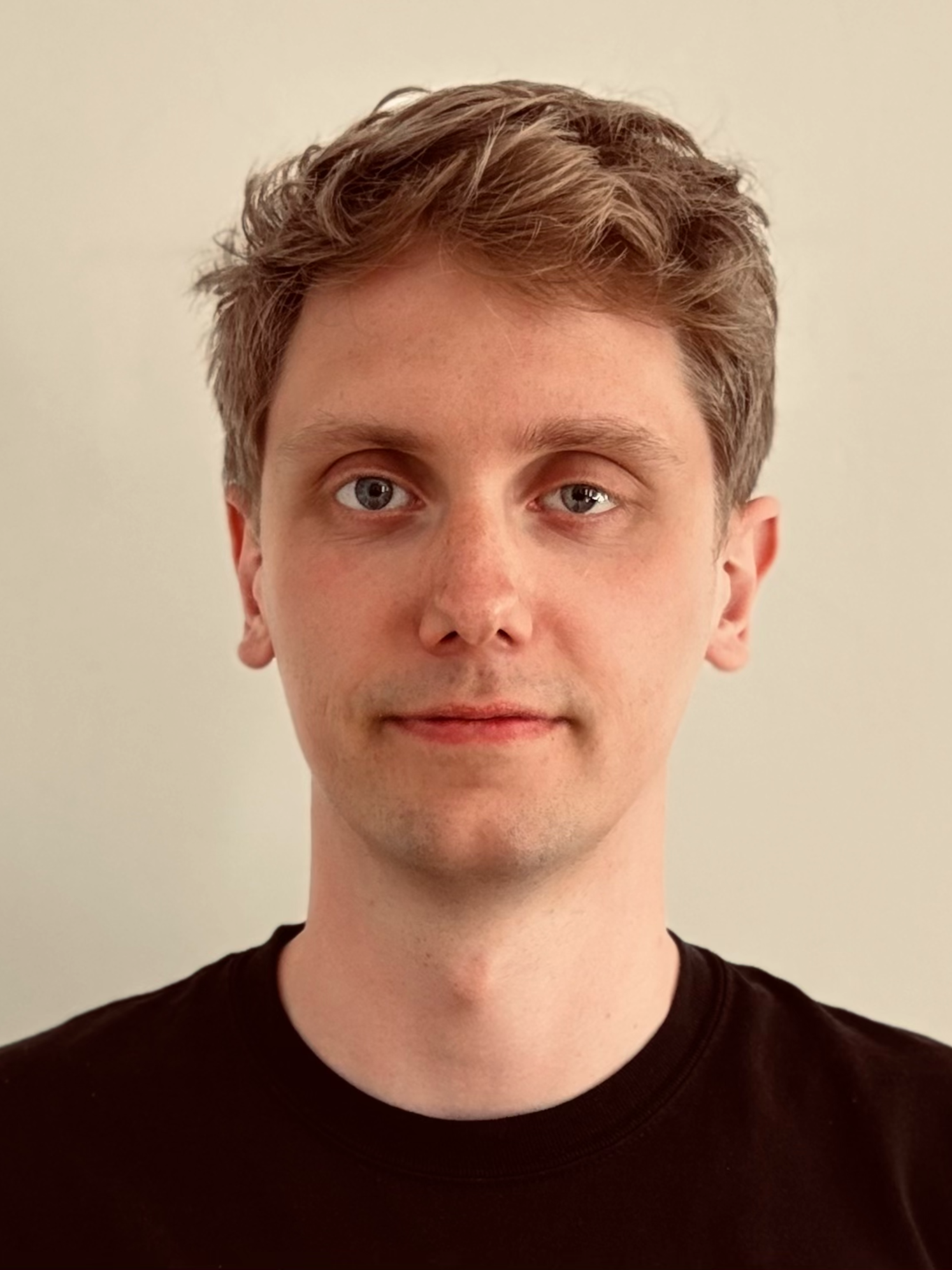}}]{Flemming Holtorf}~received the B.S. degree in mechanical engineering and the M.S. degree in process systems engineering from RWTH Aachen University, Aachen, Germany, in 2016 and 2018, respectively. He received the Ph.D. degree from MIT in 2024, where his research at the Computer Science and Artificial Intelligence Laboratory (CSAIL) focused on the intersection of optimization and control theory, scientific computing, and quantum information. He is currently an AI Research Scientist with Axiomatic\_AI. His research interests include control and optimization theory, scientific computing, and their role in automating the scientific process with AI.
\end{IEEEbiography}

\begin{IEEEbiography}[{\includegraphics[width=1in,height=1.25in,clip,keepaspectratio]{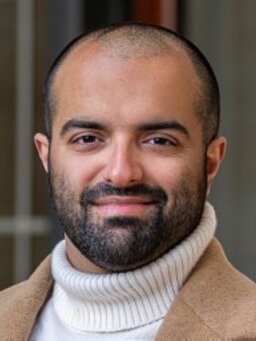}}]{Kenaish AlQubaisi}~is currently a Postdoctoral Researcher at the Massachusetts Institute of Technology (MIT). He received his Ph.D. degree in Electrical Engineering from Boston University in 2023, an M.Sc. degree in Microsystems Engineering from the Masdar Institute in 2015, and a B.Sc. degree in Electrical Engineering from Purdue University. His research interests are in silicon photonics on CMOS platforms, spanning integrated photonic devices and circuits for optical computation and quantum information processing, with an emphasis on electronic-photonic integration for machine learning hardware acceleration.
\end{IEEEbiography}

\begin{IEEEbiography}[{\includegraphics[width=1in,height=1.25in,clip,keepaspectratio]{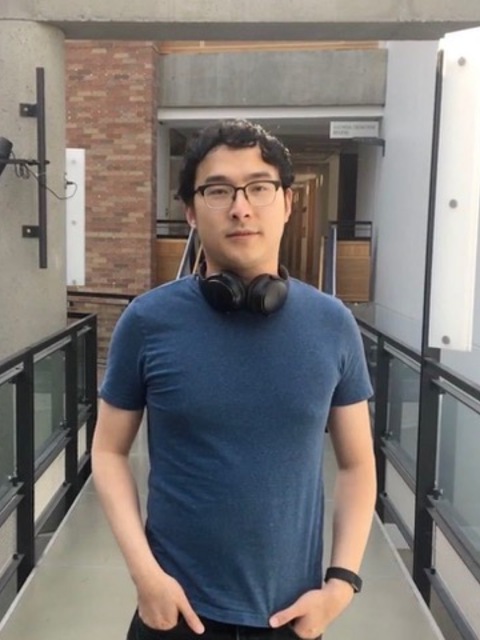}}]{Danyang Zhuo}~is an Assistant Professor of Computer Science at Duke University. He received his Ph.D. from the Paul G. Allen School of Computer Science and Engineering at the University of Washington, and B.S in Electrical Engineering from the University of Illinois Urbana-Champaign. His research interests are computer networking and machine learning systems. His research has received NSF CAREER Award, IEEE/ACM MICRO Best Paper Award, USENIX Security Distinguished Paper Award, USENIX FAST Best Paper Award, Google Academic Research Award, Meta Research Award, Amazon Research Award, and IBM Academic Award.
\end{IEEEbiography}

\begin{IEEEbiography}[{\includegraphics[width=1in,height=1.25in,clip,keepaspectratio]{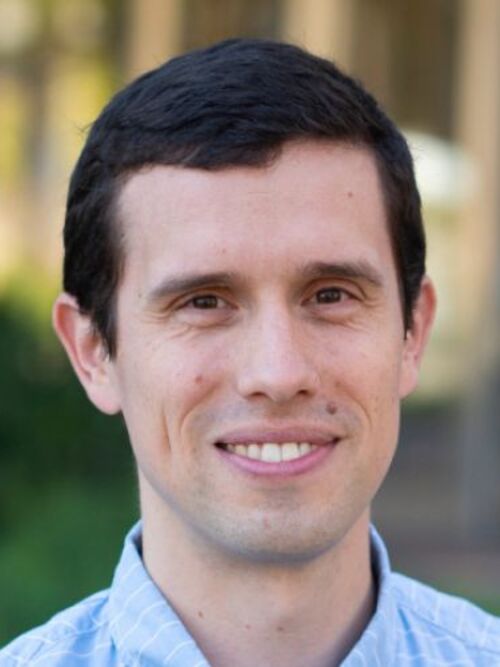}}]{Norbert M. Linke}~is an Associate Professor of Physics, a Fellow of the Joint Quantum Institute, and the Director of QLab, the National Quantum Laboratory at the University of Maryland (UMD). His research focuses on quantum applications of trapped ions and the realization of quantum algorithms and quantum network applications. Born in Munich, Germany, he graduated from the University of Ulm, and received his doctorate at the University of Oxford, UK, working on micro-fabricated ion-traps and microwave-addressing of ions. He spent four years as a postdoc and research scientist in the group of Chris Monroe at UMD where he led a project that turned a physics experiment into a programmable quantum computer. He became an assistant professor at UMD in 2019. From 2022, he spent three years at the Duke Quantum Center at Duke University in Durham, NC, and in 2025, he returned to UMD to become the Director of QLab, a user facility providing access to quantum computers for researchers, educators, and start-ups.
\end{IEEEbiography}

\begin{IEEEbiography}[{\includegraphics[width=1in,height=1.25in,clip,keepaspectratio]{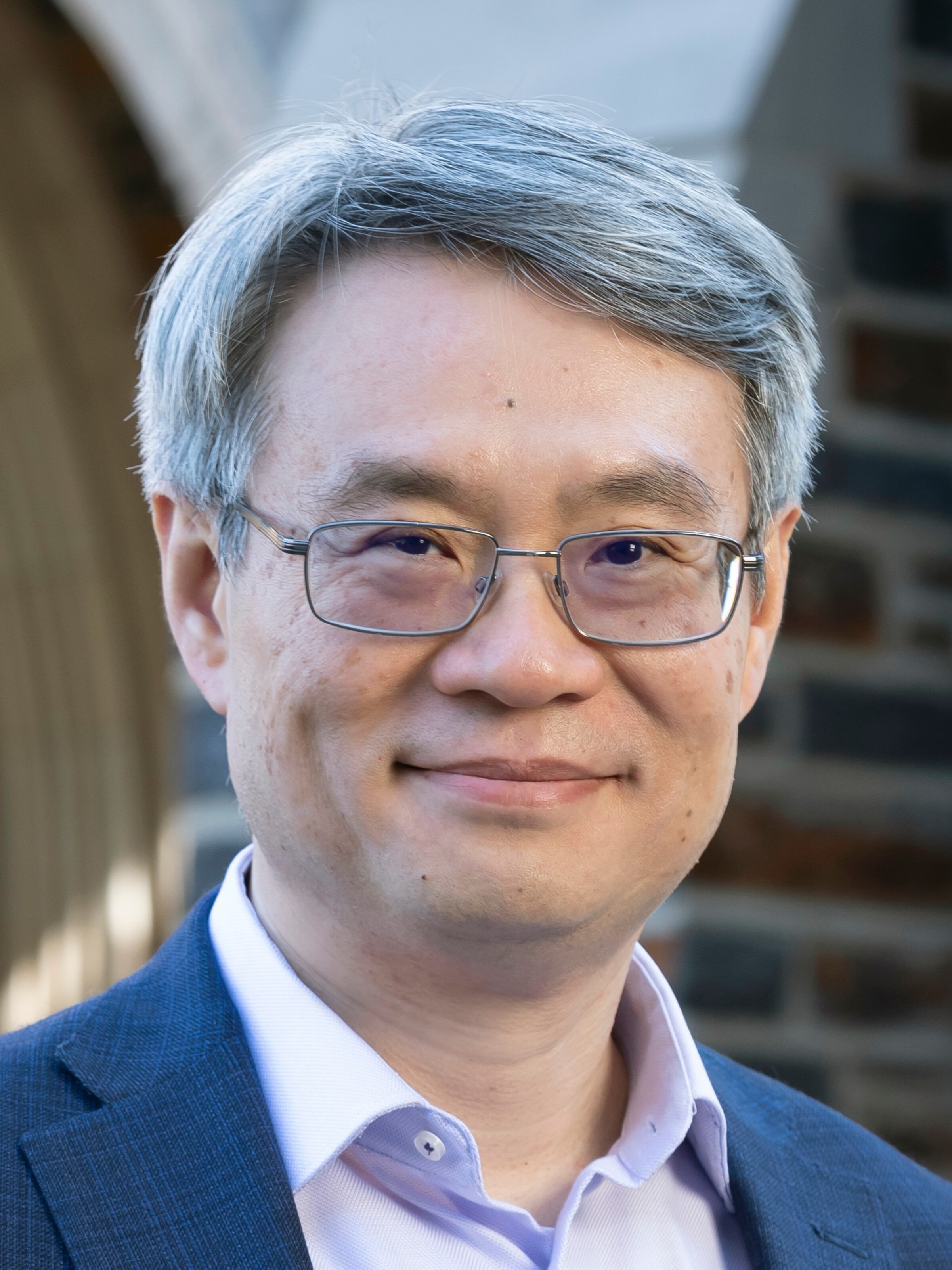}}]{Yiran Chen}~is the John Cocke Distinguished Professor of Electrical and Computer Engineering at Duke University. He serves as the Principal Investigator and Director of the NSF AI Institute for Edge Computing Leveraging Next Generation Networks (Athena). His research group focuses on innovations in emerging memory and storage systems, machine learning and neuromorphic computing, and edge computing. Dr. Chen has authored over 700 publications and holds 96 U.S. patents. His work has received widespread recognition, including two Test-of-Time Awards and 14 Best Paper/Poster Awards. He is the recipient of the IEEE Circuits and Systems Society’s Charles A. Desoer Technical Achievement Award and the IEEE Computer Society’s Edward J. McCluskey Technical Achievement Award. Dr. Chen is a Fellow of the AAAS, ACM, IEEE, and NAI, and a member of the European Academy of Sciences and Arts.
\end{IEEEbiography}

\begin{IEEEbiography}[{\includegraphics[width=1in,height=1.25in,clip,keepaspectratio]{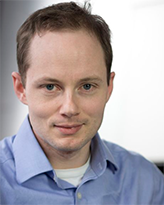}}]{Dirk Englund}~is a Professor of Electrical Engineering and Computer Science Department at MIT. 
He received the B.S. degree in physics from the California Institute of Technology, Pasadena, CA, USA, in 2002. He subsequently spent one year as a Fulbright Fellow at the Eindhoven University of Technology, Eindhoven, The Netherlands, where he designed and built a system for ultrafast magneto-optic nanoscopy. He earned the M.S. degree in electrical engineering and the Ph.D. degree in applied physics from Stanford University, Stanford, CA, USA, in 2008. From 2008 to 2010, he was a Postdoctoral Fellow at Harvard University, Cambridge, MA, USA. Prior to joining MIT, he was an Assistant Professor of Electrical Engineering and of Applied Physics at Columbia University, New York, NY, USA.
He has been recognized with the DARPA Young Faculty Award (2012), the IBM Faculty Award (2012), the Presidential Early Career Award for Scientists and Engineers (2011), the Sloan Research Fellowship in Physics (2011), the IEEE-Eta Kappa Nu Outstanding Young Professional Award (2012), and an Intelligence Community (IC) Postdoctoral Fellowship (2008).
\end{IEEEbiography}

\begin{IEEEbiography}[{\includegraphics[width=1in,height=1.25in,clip,keepaspectratio]{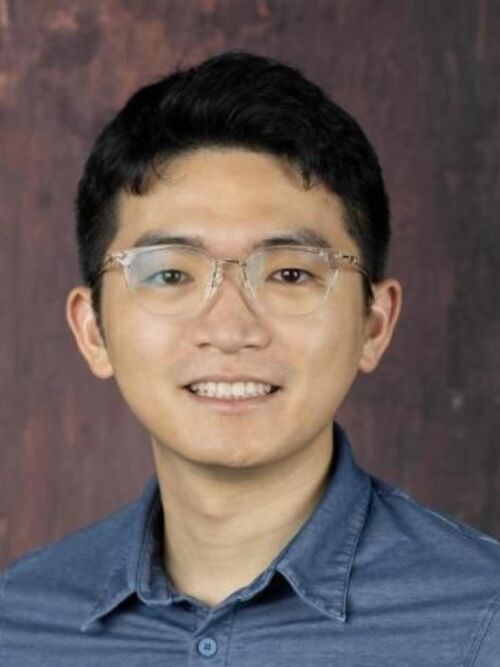}}]{Tingjun Chen}~is the Nortel Networks Assistant Professor of Electrical and Computer Engineering at Duke University.
He received the B.Eng. degree in electronic engineering from Tsinghua University in 2014, the Ph.D. degree in electrical engineering from Columbia University in 2020, and was a Postdoctoral Associate with Yale University from 2020 to 2021.
His research focuses on next-generation wireless, mobile, and optical networked systems, bridging theoretical foundations and experimental platforms.
He has received multiple awards, including the NSF CAREER Award, Google Research Scholar Award, IBM Academic Awards, and Columbia Engineering Morton B. Friedman Memorial Prize for Excellence.
He is also a co-recipient of several paper awards from ACM CoNEXT, ACM MobiHoc, IEEE MTT-S IMS, IEEE/Optica OFC, and ECOC.
His Ph.D. thesis was recognized by the ACM SIGMOBILE Dissertation Award Runner-up.
\end{IEEEbiography}


\EOD

\end{document}